\documentclass[conference]{IEEEtran}
\IEEEoverridecommandlockouts
\usepackage[table,xcdraw]{xcolor}
\usepackage{amsmath,amssymb,amsfonts}
\usepackage[ruled, vlined, linesnumbered]{algorithm2e}
\usepackage{multirow}
\usepackage{graphicx}
\usepackage{booktabs} 
\usepackage{pifont}%
\usepackage{cite}
\PassOptionsToPackage{hyphens}{url}\usepackage{hyperref}
\usepackage{algorithmic}
\usepackage{textcomp}
\usepackage{tikz}
\usepackage{pgfplots}
\usetikzlibrary{patterns}
\usepgfplotslibrary{groupplots}
\usepackage{subfig}
\usepackage[export]{adjustbox}
\usepackage{enumitem}
\usepackage[normalem]{ulem}
\usepackage{colortbl}
\usepackage{array}
\usepackage{tabu}
\usepackage{xspace,balance,tabularx}
\newenvironment{customlegend}[1][]{%
    \begingroup
    \csname pgfplots@init@cleared@structures\endcsname
    \pgfplotsset{#1}%
}{%
    \csname pgfplots@createlegend\endcsname
    \endgroup
}%
\def\addlegendimage{\csname pgfplots@addlegendimage\endcsname}

\newtheorem{theorem}{Theorem}
\newtheorem{definition}{Definition}
\newtheorem{corollary}{Corollary}[theorem]
\newtheorem{example}{Example}
     
\DeclareMathOperator*{\argmin}{argmin}
\DeclareMathOperator{\vol}{Vol}
\DeclareMathOperator{\cut}{Cut}

\newcommand{\omac}{\texttt{O2MAC}\xspace}

\newcommand{\hdmi}{\texttt{HDMI}\xspace}
\newcommand{\dmg}{\texttt{DMG}\xspace}

\newcommand{\uramn}{\texttt{URAMN}\xspace}

\newcommand{\mcgc}{\texttt{MCGC}\xspace}
\newcommand{\mvagc}{\texttt{MvAGC}\xspace}
\newcommand{\magc}{\texttt{MAGC}\xspace}

\newcommand{\lmgec}{\texttt{LMGEC}\xspace}
\newcommand{\twocmv}{\texttt{2CMV}\xspace}
\newcommand{\mega}{\texttt{MEGA}\xspace}
\newcommand{\conn}{\texttt{CONN}\xspace}

\newcommand{\pane}{\texttt{PANE}\xspace}
\newcommand{\aneci}{\texttt{AnECI}\xspace}
\newcommand{\wmsc}{\texttt{WMSC}\xspace}
\newcommand{\magcn}{\texttt{MAGCN}\xspace}

\newcommand{\main}{\texttt{SGLA}\xspace}
\newcommand{\ours}{\texttt{SGLA}\xspace} %
\newcommand{\qours}{\texttt{SGLA}+\xspace}%
\newcommand{\mvag}{MVAG\xspace}
\newcommand{\mvags}{MVAGs\xspace}

\newcommand{\netmf}{\texttt{NetMF}\xspace}
\newcommand{\sketchne}{\texttt{SketchNE}\xspace}

\newcommand{\cobyla}{\text{Cobyla}\xspace}

\newcommand{\RN}{\mathbb{R}\xspace}

\newcommand{\GS}{\mathcal{G}\xspace}
\newcommand{\VS}{\mathcal{V}\xspace}
\newcommand{\CS}{\mathcal{C}\xspace}
\newcommand{\ES}{\mathcal{E}\xspace}
\newcommand{\LS}{\mathcal{L}\xspace}

\newcommand{\cons}{{\Omega}\xspace}
\newcommand{\XM}{\mathbf{X}\xspace}
\newcommand{\AM}{\mathbf{A}\xspace}
\newcommand{\LM}{\mathbf{L}\xspace}
\newcommand{\DM}{\mathbf{D}\xspace}
\newcommand{\IM}{\mathbf{I}\xspace}

\newcommand{\ThM}{\boldsymbol{\Theta}\xspace}

\newcommand{\wv}{\mathbf{w}\xspace}

\newcommand{\nv}{r\xspace}
\newcommand{\obj}{h\xspace}
\newcommand{\fun}{h\xspace}
\newcommand{\qreg}{{h}_{\ThM}\xspace}
\newcommand{\qopt}{{h}_{\ThM^*}\xspace}

\newcommand{\tol}{\epsilon\xspace}
\newcommand{\tmax}{T_\text{max}\xspace}
\newcommand{\ralpha}{\alpha_r\xspace}

\newcommand{\transpose}{^\mathsf{T}}
\newcommand{\rvect}[1]{\begin{bmatrix} #1 \end{bmatrix}}

\newcommand{\eat}[1]{{}}

\newcommand*{\rowstyle}[1]{%
  \gdef\@rowstyle{#1}%
  \@rowstyle\ignorespaces%
}

\newcolumntype{=}{%
  >{\gdef\@rowstyle{}}%
}

\newcolumntype{+}{%
  >{\@rowstyle}%
}

\newcommand{\revision}[1]{{#1}}

\renewcommand{\eqref}[1]{Eq. (\ref{#1})}

\newcommand{\first}{\cellcolor{blue!30}}
\newcommand{\second}{\cellcolor{blue!20}} %
\newcommand{\third}{\cellcolor{blue!8}}

\newcommand{\ie}{{\it i.e.},\xspace}
\newcommand{\eg}{{\it e.g.},\xspace}

\newcommand{\stitle}[1]{\vspace*{0.6mm}\noindent{\bf #1.\/}}

\def\BibTeX{{\rm B\kern-.05em{\sc i\kern-.025em b}\kern-.08em
    T\kern-.1667em\lower.7ex\hbox{E}\kern-.125emX}}

\begin{document}

\title{Efficient Integration of Multi-View Attributed Graphs for
Clustering and Embedding}

\author{\IEEEauthorblockN{Yiran Li}
\IEEEauthorblockA{
\textit{The Hong Kong Polytechnic University}\\
yi-ran.li@connect.polyu.hk}
\and
\IEEEauthorblockN{Gongyao Guo}
\IEEEauthorblockA{
\textit{The Hong Kong Polytechnic University}\\
gongyao.guo@connect.polyu.hk}
\and
\IEEEauthorblockN{Jieming Shi}
\IEEEauthorblockA{\textit{Hong Kong Polytechnic University} \\
jieming.shi@polyu.edu.hk}
\and
\IEEEauthorblockN{Sibo Wang}
\IEEEauthorblockA{
\textit{The Chinese University of Hong Kong}\\
swang@se.cuhk.edu.hk}
\and
\IEEEauthorblockN{Qing Li}
\IEEEauthorblockA{\textit{The Hong Kong Polytechnic University} \\
csqli@comp.polyu.edu.hk}
}

\maketitle

\begin{abstract}
A multi-view attributed graph (\mvag) $\GS$ captures the diverse relationships and properties of real-world entities through multiple graph views and attribute views.
Effectively utilizing all views in $\GS$ is essential for \mvag clustering and embedding, which are important for applications like recommendation systems, anomaly detection, social network analysis,  etc. 
Existing methods either achieve inferior result quality or incur significant computational costs to handle large-scale \mvags. 

In this paper,  we present a spectrum-guided Laplacian aggregation scheme with an effective objective formulation and two efficient algorithms \ours and \qours, to cohesively integrate all views of $\GS$ into an \mvag Laplacian matrix, which readily enables classic graph algorithms to handle $\GS$ with superior performance in clustering and embedding tasks.
\revision{We begin by conducting a theoretical analysis to design an integrated objective that consists of two components, the eigengap and connectivity objectives, aiming to link the spectral properties of the aggregated \mvag Laplacian  with  the underlying community and connectivity properties of $\GS$.} 
A constrained  optimization problem is then formulated for the integration, which is computationally expensive to solve. Thus, we first develop the \ours algorithm, which already achieves excellent performance compared with existing methods.
To further enhance efficiency, we design \qours to reduce the number of costly objective evaluations via sampling and approximation to quickly find an approximate optimum. 
Extensive experiments compare our methods against 12 baselines for clustering and 8 baselines for embedding on  8 multi-view attributed graphs, validating the superior performance of \ours and \qours in terms of result quality and efficiency.
Compared with the most effective baselines, our methods are significantly faster, often by up 
 to orders of magnitude. 
Our implementation is available at \url{https://github.com/CyanideCentral/SGLA/}.

\end{abstract}


\section{Introduction}

\begin{figure}[!t]

\centering

\includegraphics[width=0.86\columnwidth]{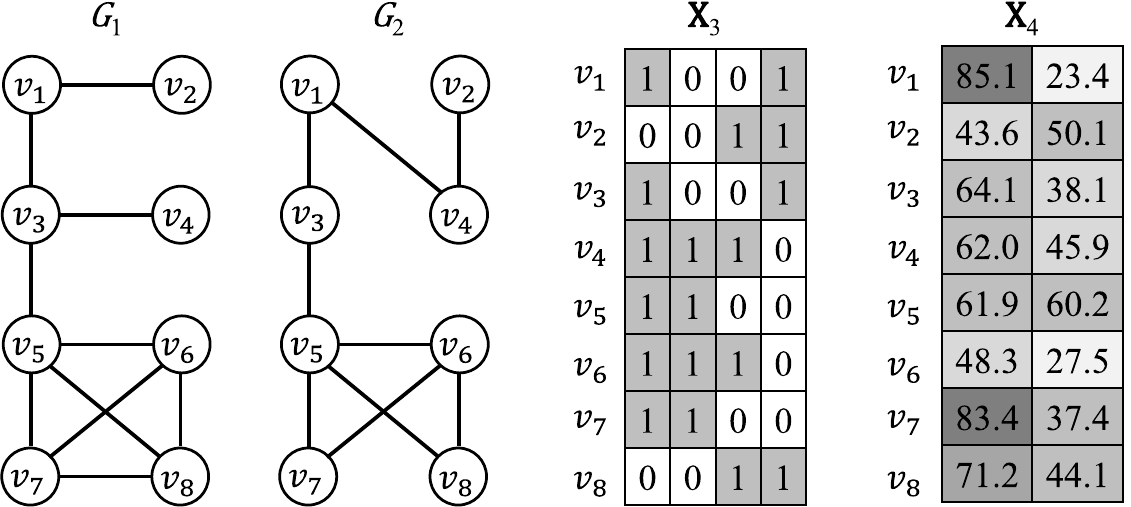}
\vspace{-1mm}
\caption{\revision{Multi-view attributed graph $\mathcal{G}$ with two graph views $G_1$ and $G_2$, and two attribute views $\XM_3$ and $\XM_4$ of categorical and numerical attributes respectively.}} \label{fig:data-model}
\vspace{-4mm}
\end{figure}

A \textit{multi-view attributed graph}  (\mvag) describes a set of entities with multiple \textit{graph views} and \textit{attribute views}, illustrating their relationships and properties from various perspectives or data sources. For example, regarding a group of people, one graph view may focus on their social relations on Facebook, whereas another graph view 
may represent their business connections on LinkedIn. 
Moreover, attribute views may comprise diverse numerical, categorical, or visual features. 
\revision{Graph analytics for \mvags, especially clustering and embedding, are of particular interest as they find important applications. For instance, clustering on \mvags constructed from  visual descriptors is effective for neuroimaging analysis of diseases \cite{zhangMultiViewGraphConvolutional2018}.  MVAG embeddings  are useful in recommendation systems in e-commerce \cite{wangM2GRLMultitaskMultiview2020}, spam detection on social networks \cite{liSSDMVSemiSupervisedDeep2018}, and predicting drug-disease associations in bioinformatics \cite{fuMVGCNDataIntegration2022}.}
\revision{Figure \ref{fig:data-model} presents a minimal example of an \mvag of 8 entities, described by 2 graph views and 2 attribute views.}

It is crucial but challenging to manage complex MVAGs to holistically utilize the graph views and attribute views for the clustering and embedding tasks, particularly with large MVAGs. In an MVAG $\GS$, different graph views can display varying topological structures, and the graph and attribute views represent distinct data models that cannot be directly integrated. Moreover, the considerable data volumes typical in real-world applications pose a significant challenge to efficiency and scalability. These challenges hinder the effective management of MVAG data in an efficient manner.

As reviewed in Section \ref{sec:related}, an array of approaches  \cite{xu2012model, fanOne2MultiGraphAutoencoder2020, liu2022robust, Zhang2022UnsupervisedRL, yangPANEScalableEffective2023} are only designed for attributed graphs with one graph view or one attribute view. Other methods~\cite{zongWeightedMultiViewSpectral2018,fangComprehensiveSurveyMultiview2023} handle attribute views without considering graphs, and thus they often yield suboptimal results for \mvags in the experiments.
Existing methods specialized for \mvag clustering or embedding are often built upon sophisticated graph neural network operations~\cite{chengMultiviewAttributeGraph2021, Mo2023DisentangledMG}, and consequently struggle with efficiency and scalability. Several clustering methods ~\cite{panMultiviewContrastiveGraph2021,linMultiViewAttributedGraph2023} attempt to learn a graph structure that aligns with all views in $\GS$, requiring a huge number of variables to be solved.
Summing up, existing methods either produce subpar results or require excessive computational resources to manage large-scale MVAGs.

In this paper, we focus on the important problem of how to utilize the rich semantics of all views in an \mvag $\GS$, and develop an effective and efficient \underline{S}pectrum-\underline{G}uided \underline{L}aplacian \underline{A}ggregation approach, exploiting the intrinsic spectral properties to cohesively integrate all views of $\GS$ into an \mvag Laplacian matrix $\LS$ for clustering and embedding. 

Our main designs include a carefully formulated objective for the integration (Section \ref{sec:sgfobjective}), an efficient method \ours producing high-quality results (Section \ref{sec:basemethod}), and \qours that boosts efficiency further while maintaining the effectiveness (Section \ref{sec:pi}).
Specifically,  our strategy is to perform a weighted aggregation of normalized Laplacian matrices from all views in $\GS$ to produce the integrated Laplacian $\LS$. 
However, a critical challenge is choosing appropriate view weights to produce an effective $\LS$ that preserves the fundamental characteristics of $\GS$, \ie community structure and node connectivity, which are important for clustering and embedding. 
With this in mind, we design an objective function on the basis of spectral graph theory.
In particular, we align the spectrum of $\LS$ with a normalized-cut community measure and a graph conductance measure, and propose eigengap and connectivity objectives accordingly. 
The overall objective combines  the eigengap and connectivity objectives, assisted with a regularization term, to determine the appropriate weights of each view in $\GS$ to get $\LS$. 
It is challenging to optimize the overall objective in search of appropriate view weights, since it is infeasible to exhaust all weight combinations, and the objective evaluation is computationally expensive. To mitigate the challenges, 
in Section \ref{sec:algorithms}, we develop the \ours algorithm to find a desirable solution, which already achieves excellent performance compared with existing methods. Nevertheless, \ours needs to evaluate the objective at every iteration, causing significant overhead. To boost efficiency with fewer objective evaluations, we develop the \qours algorithm, which employs sampling and  interpolation approximation to quickly find an effective solution. 
In our experiments, we couple \ours and \qours with spectral clustering and embedding methods, to compare them against 12 clustering baselines and 8 embedding baselines over 8 real-world \mvag datasets. \ours and \qours achieve better performance in terms of both effectiveness and efficiency. For instance, on large-scale datasets, such as MAG-phy with 4 views and 2.35 million nodes, our methods efficiently produce high-quality results, while most baselines fail to scale.

The contributions of this paper are summarized as follows:
\begin{itemize}[leftmargin=*]
    \item We propose an efficient and effective spectrum-guided aggregation scheme for \mvag clustering and embedding. 
    \item \revision{We derive a novel objective formulation, consisting of the  eigengap and connectivity objectives, to find appropriate view weights that preserve the community structure and node connectivity in \mvags.}
    \item We develop two efficient algorithms, \ours and \qours, with several speedup techniques to optimize the objective.

    \item  Extensive experiments on 8 real-world \mvags validate the superior performance  of \ours and \qours.
\end{itemize}

The rest of the paper is organized as follows. 
{Section \ref{sec:related} reviews related work. Section \ref{sec:prelim} illustrates the preliminaries of \mvags and presents the problem statement. To solve the problem, in Section \ref{sec:sgfobjective}, we develop the full objective formulation with the proposed eigengap and connectivity objectives.} Then, in Section \ref{sec:algorithms}, we design two efficient algorithms  \ours and \qours to solve the objective. Experiments are reported in  Section \ref{sec:experiments}, and Section \ref{sec:conclusion} concludes the paper.

\section{Related Work}\label{sec:related}

This section reviews the related work on  multi-view attributed graphs and similar data models. 

{For basic attributed graphs with one graph and one attribute view, 
attributed network embedding and clustering have been extensively studied in the literature \cite{zhouGraphClusteringBased2009a,xu2012model,mrabahRethinkingGraphAutoEncoder2023,yang2020scaling, liu2022robust, yangPANEScalableEffective2023}. 
For instance, Bayesian probabilistic model \cite{xu2012model} and graph auto-encoder \cite{mrabahRethinkingGraphAutoEncoder2023} have been adopted for clustering.}
An attributed graph embedding algorithm~\cite{yang2020scaling} captures the multi-hop affinity between nodes and attributes~\cite{yang2020scaling,yangPANEScalableEffective2023}. 
{GNN-based embedding method \texttt{AnECI}~\cite{liu2022robust} strengthens the robustness by preserving communities, while \texttt{CONN}~\cite{tanCollaborativeGraphNeural2024} adopts selective graph diffusion with attribute augmentation.} 
\revision{In \cite{LinLGXY23}, the authors propose using diverse pretext tasks to capture different signals in graphs with heterophily into embeddings.} 
These approaches do not consider the unique characteristics of \mvags and tend to yield suboptimal performance.

On \mvags, \mcgc~\cite{panMultiviewContrastiveGraph2021} and \magc~\cite{linMultiViewAttributedGraph2023} construct a consensus graph matrix for \mvag clustering in $O(n^2)$ time where $n$ is the number of nodes, by optimizing a dense $n\times n$ adjacency matrix that minimizes reconstruction loss on each view.  
\mvagc~\cite{linGraphFilterbasedMultiview2021a} improves their complexity to linear time with node sampling while compromising result quality and stability. Their problem formulations neglect the overall structure of $\GS$ and suffer from the difficulty of optimizing at least $O(n)$ variables.
\revision{Besides, various GNNs have been adopted for \mvags, including \omac~\cite{fanOne2MultiGraphAutoencoder2020}, \texttt{MAGCN}~\cite{chengMultiviewAttributeGraph2021} and \dmg~\cite{Mo2023DisentangledMG}. For instance, the clustering model \magcn uses graph auto-encoders to map each view to latent representations for reconstruction.  \cite{KhanB19} combines graph views by fusing Laplacian matrices and trains a semi-supervised GNN.}
Other studies~\cite{parkUnsupervisedAttributedMultiplex2020, jingHDMIHighorderDeep2021, Zhang2022UnsupervisedRL} adopt mutual information models to learn view-specific node embeddings and aggregate them with attention mechanism. These methods incur high costs of training and exhibit inferior performance on 
\mvags. 
\texttt{MEGA}~\cite{whang2020mega} tackles semi-supervised \mvag clustering by joint nonnegative matrix factorization. 
\twocmv~\cite{luong2020novel} learns the consensus and complementary components from each view via matrix factorization with $O(n^2)$ complexity. \lmgec~\cite{fettalSimultaneousLinearMultiview2023} addresses clustering and embedding within a unified formulation, while the embedding quality is inferior to its clustering performance.

There are also algorithms that only handle attribute views for clustering, as surveyed in~\cite{fangComprehensiveSurveyMultiview2023}. For instance, a work~\cite{zongWeightedMultiViewSpectral2018} adopts a weighting objective to minimize the subspace distances between its integration result and each view. \revision{A recent study~\cite{yacobiSpecRaGERobustGeneralizable2024} learns a robust fused representation of noisy attributes via meta-learning. However, these methods \cite{zongWeightedMultiViewSpectral2018,khanApproximateGraphLaplacians2021,yacobiSpecRaGERobustGeneralizable2024}  do not consider graph topological properties of \mvags.}

\section{Preliminaries and Problem Statement}\label{sec:prelim}
\label{sec:prelimAndStatement}

\subsection{Preliminaries}

\noindent
\textbf{Multi-view Attributed Graph.}
\revision{Figure \ref{fig:data-model} shows an MVAG of 8 nodes with 4 views, including 2 graph views and 2 attribute views. Graph views $G_1$ and $G_2$ are two simple graphs, while attribute views $\XM_3$ and $\XM_4$ are binary and numerical attributes, respectively.}
Formally, we denote an MVAG $\GS$ with $p$ graph views and $q$ attribute views by $\GS = \{\VS, \ES_1, \dots, \ES_p,\allowbreak \XM_{p+1}, \dots, \XM_{p+q}\}$, where $\VS$ is the set of $n$ nodes, $\ES_i$ is the set of edges in the $i$-th graph view $G_i = \{\VS, \ES_i\}$ that is a simple graph, and matrix $\XM_{p+j}$ contains the values in the $j$-th attribute view. 
We focus on \mvags with a total of $r=p+q>2$ views. Denote the number of edges in  $G_i$ by $m_i$, and the total number of edges in $\GS$ as $m=\sum_{i=1}^p m_i$. $G_i$ has an adjacency matrix $\AM_i\in \RN^{n\times n}$, and a node $v_a$ in $G_i$ has a generalized degree $\delta_i(v_a)$ equal to the total weight of incident edges.

\stitle{Normalized Laplacian}
\revision{The normalized Laplacian  of a simple graph $G$ is  $\LM(G)=\IM_n - \DM^{-\frac{1}{2}}\AM\DM^{-\frac{1}{2}}$, where $\DM$ is the diagonal degree matrix and $\IM_n$ is the identity matrix \cite{chungSpectralGraphTheory1997, spielmanSpectralGraphTheory2007}.}

\stitle{\mvag Clustering and Embedding} The two analytic tasks for \mvags are stated as follows:
\begin{itemize}[leftmargin=*]
\item \textit{Clustering} is to divide the 
nodes in $\GS$ into $k$ disjoint non-empty subsets $\{\CS_1, \dots, \CS_k\}$, \ie $k$ clusters, such that nodes within each cluster tend to form dense connections in graph views and share similar values in attribute views. 
\item \textit{Embedding} is to map each node $\GS$ to a low-dimensional embedding vector that captures its features inherent to the graph views and attribute views in $\GS$. 
\end{itemize}

Table \ref{tbl:notations} lists the frequently used notations.

\subsection{Problem Statement}\label{sec:problem}
Given an \mvag $\GS$, our goal is to generate an \mvag Laplacian matrix $\LS$ as the multi-view integration, which empowers classic methods to handle clustering and embedding tasks on $\GS$. Previous approaches that construct a new graph from scratch typically require at least $O(n)$ variables to be determined~\cite{panMultiviewContrastiveGraph2021,linMultiViewAttributedGraph2023}, which is computationally expensive. Contrarily, we adopt an intuitive yet effective weighted aggregation from the Laplacian matrices of all views into the \mvag Laplacian $\LS$, where only $r$ variables need to be optimized. In what follows, we describe the problem statement of multi-view attributed graph integration, while the objective function to solve the problem is formally developed in Section \ref{sec:sgfobjective}.

\stitle{View Laplacians} \revision{Let $\LM_i$ denote the Laplacian of the $i$-th view of $\GS$, called the $i$-th view Laplacian. If this view is a graph view $G_i$, $\LM_i$ is its normalized Laplacian $\LM(G_i)$.
If it is an attributed view $\XM_i$, we adopt a prevalent way~\cite{luong2020novel} to construct a $K$-nearest neighbor (KNN) graph $G_K(\XM_i)$, consequently deriving its Laplacian $\LM_i=\LM(G_K(\XM_i))$.
In this KNN graph, every node is connected to $K$ neighbors with the highest attribute similarity measured by cosine similarity, and each edge is weighted by attribute similarity.} %

\begin{table}[!t]
\centering
\renewcommand{\arraystretch}{1.1}
\begin{footnotesize}
\caption{Frequently used notations.}\vspace{-1mm} \label{tbl:notations}
\resizebox{\columnwidth}{!}{
	\begin{tabular}{|p{0.90in}|p{2.15in}|}
		\hline
		{\bf Notation} &  {\bf Description}\\
		\hline
		\!$\GS=\{\VS,\ES_1, \dots,\allowbreak \ES_p, \XM_{p+1}, \dots, \XM_{r}\}$\!  & $\GS$ is an \mvag with node set $\VS$ and $r=p+q$ views, including $p$ graph views with edge sets $\ES_1, \dots, \ES_p$, and  $q$ attribute views $\XM_{p+1}, \dots, \XM_{r}$.\\
		\hline
		$n$   & The number of nodes in $\GS$.\\
            \hline
            $m_i$, $m$   & The number of edges in the $i$-th graph view, and the total number of edges in all graph views.\\
            \hline
            $k$   & The number of clusters or classes in $\GS$.\\
            \hline
        $G_K(\XM_j)$   &  The $K$-nearest neighbor graph constructed from the attribute view $\XM_j$.  \\
	       \hline
        $\LM(G)$   & The normalized Laplacian of a simple graph $G$. \\
	       \hline
    		$\LM_i$   & The $i$-th view Laplacian of $\GS$. \\
    	    \hline
            $\LS$   & The \mvag Laplacian of $\GS$, defined in \eqref{eq:mvlap}. \\
	    \hline
            $\wv=[w_1,\dots,w_r]$   & A weight vector for $r$ view Laplacians. \\
            \hline
            $\lambda_i$   & The $i$-th smallest eigenvalue of $\LS$. \\
            \hline
            $g_k(\LS)$   & The eigengap objective. \\
            \hline
            $\lambda_2(\LS)$   & The connectivity objective. \\
            \hline
            $\fun(\wv)$   & The spectrum-guided objective function. \\
            \hline
            $\wv^*$   & The weights minimizing $\fun$ found by \ours. \\
            \hline
            $\qreg(\wv)$, $\qopt(\wv)$   & An interpolation of $\fun$ with coefficients $\ThM$, or the optimal coefficients $\ThM^*$. \\
            \hline
            $\wv_0,\dots,\wv_r$   & $r+1$ weight vectors sampled for interpolation. \\
            \hline
            $\wv^\dagger$   & The weights minimizing $\qopt$ found by \qours. \\
            \hline
	\end{tabular}
}
\end{footnotesize}
\vspace{-2mm}
\end{table}

\stitle{\mvag Laplacian}
Intuitively,  each view in $\GS$ can complement each other with its own information, and thus effective integration is vital to MVAG clustering and embedding. We define the \mvag Laplacian $\LS$ as a weighted aggregation of all view Laplacians, where $w_i$ is the $i$-th view weight.
\begin{equation}\label{eq:mvlap}
    \LS=\sum\limits_{i=1}^{\nv} w_i \LM_i, \textrm{ where } \sum\limits_{i=1}^{\nv} w_i =1 \revision{\text{ and any } w_i\geq 0.}
\end{equation}

{As a weighted combination of graph structures and attribute similarities across all views, this $\LS$ can be interpreted as one integrated view of the \mvag and used for downstream tasks.}
For \mvag clustering, we employ the spectral clustering method in~\cite{yuMulticlassSpectralClustering2003} using the bottom eigenvectors of $\mathcal{L}$ to assign clusters. 
We utilize $\LS$ as the input for graph embedding methods based on matrix factorization~\cite{qiuNetworkEmbeddingMatrix2018a,xieSketchNEEmbeddingBillionScale2023} to enable them for \mvag embedding.

\stitle{Problem Statement} The quality of $\LS$ is crucial for empowering classic spectral clustering and network embedding methods to outperform state-of-the-art dedicated methods, while $\LS$ solely depends on the $r$ view weights in \eqref{eq:mvlap}. Thus, our main research problem is to decide on proper view weights to produce an \mvag Laplacian for effective clustering and embedding quality.

\section{SGLA Objective} \label{sec:sgfobjective}

\stitle{Objective Overview} 
To assign the view weights for \mvag integration, trivial solutions such as utilizing a single view or allocating weights uniformly both compromise the performance, as validated in our experiments.
\revision{Intuitively, we should cohesively leverage all views in the \mvag to produce $\LS$, recognizing that different views may contribute variably.} %
\revision{Community structures and connectivity properties are fundamental characteristics in real-world network data~\cite{fangEffectiveCommunitySearch2016a,nagamochi2008algorithmic}, which are important for various problems, including clustering and embedding.
To preserve these underlying properties, we analyze and design  two objectives--\textit{eigengap} and \textit{connectivity}--which are combined in the full objective to produce the desired MVAG Laplacian $\LS$.} 

In Section  \ref{sec:eigengapobj}, 
we analyze and  align the spectrum of $\LS$ with the community property measured by normalized cut and propose an \textit{eigengap objective}.
In Section \ref{sec:connectivityobj}, we link the connectivity property, measured by conductance, to the spectrum of $\LS$ and devise a \textit{connectivity objective}.
In Section \ref{sec:overall-obj}, we combine the two objectives with an auxiliary regularization term to get the overall objective. The objective is formulated as a constrained nonlinear optimization,  to find the desired view weights to compute $\LS$. Figure \ref{fig:obj-example} is a running example to intuitively explain the objectives in these sections.

\subsection{Eigengap Objective} \label{sec:eigengapobj} 
\revision{In this section, we aim to build the connection between the eigenvalues  of the proposed \mvag Laplacian $\LS$  and clustering quality that is measured by normalized cut in Definition \ref{def:normcut}.} 
The normalized cut $\phi(\CS)$ of a cluster $\CS$ in a simple graph $G$ is the total weight of all outgoing edges from nodes within $\CS$ to nodes outside $\CS$ divided by the sum of degrees of all nodes in $\CS$.
A small normalized cut $\phi(\CS)$ indicates better cluster quality.

\begin{definition}\label{def:normcut}
In a graph $G$, a cluster of nodes $\CS \subset \VS$ has volume $\vol(\CS)=\sum_{v_a\in \CS}\delta(v_a)$, and $\cut(\CS)=\sum_{v_a\in \CS, v_b\notin \CS}\AM[a,b]$ is its cut value that measures the total weight of outgoing edges from nodes within $\CS$. The normalized cut of $\CS$ is defined as $\phi(\CS)=\frac{\cut(\CS)}{\vol(\CS)}$.
\end{definition}

\revision{The high-level intuition is that if a simple graph $G$ is perfectly clustered into $k$ connected components, its normalized Laplacian forms a block diagonal matrix with zero-valued eigenvalues of multiplicity $k$, \ie $0=\lambda_1=\dots=\lambda_k<\lambda_{k+1}$.  According to matrix perturbation theory~\cite{katoPerturbationTheoryLinear2013}, a small perturbation of this Laplacian should keep $\lambda_1, \dots, \lambda_{k}$  close to zero.
Consequently, a graph with well-formed $k$ clusters, \ie communities, should have small, near-zero eigenvalues $\lambda_1,...,\lambda_k$, while eigenvalue $\lambda_{k+1}$ is relatively larger, indicating \textit{a significant multiplicative eigenvalue gap} between $\lambda_{k+1}$ and $\lambda_{k}$.
The following higher-order Cheeger's inequality from~\cite{leeMultiwaySpectralPartitioning2014} provides an upper limit for $\phi(\CS)$. Using Theorem \ref{theorem:eigengap}, we can establish a link between the eigenvalue gap and high-quality clusters, as shown in Corollary \ref{corollary:eigengap}, by setting $\xi=\frac{1}{k}$.}

\begin{theorem}\label{theorem:eigengap} %
    There is a constant $c > 0$ such that for any weighted graph $G$ and $k\in \mathbb{N}$, the following holds. Let $\xi \in (0, \frac{1}{3})$ be such that $\xi k$ is an integer. If $\lambda_{(1+\xi)k}>c\frac{(\log k)^2}{\xi^9}\lambda_k$, there are at least $s\geq (1 - 3\xi)k$ nonempty disjoint sets of nodes $\CS_1, \CS_2,\dots, \CS_s \subset \VS$ such that $\phi(\CS_i)\leq O(\sqrt{\frac{\lambda_k}{\xi^3}}),\ \forall 1\leq i \leq s$.
\end{theorem}

\begin{corollary}\label{corollary:eigengap} 
    There is a constant $c > 0$ such that for any weighted graph $G$ and $k\in \mathbb{N}$, the following holds. 
    If $\lambda_{k}<\frac{1}{ck^9(\log k)^2}\lambda_{k+1}$, there are at least $s\geq k-3$ nonempty disjoint clusters $\CS_1, \CS_2,\dots, \CS_s \subset \VS$ such that the normalized cut $\phi(\CS_i)\leq O(\sqrt{k^3\lambda_k}),\ \forall 1\leq i \leq s$. %
\end{corollary}

Corollary \ref{corollary:eigengap} indicates that if the \textit{multiplicative eigenvalue gap} between $\lambda_k$ and $\lambda_{k+1}$ is large enough, the normalized cut $\phi(\CS_i)$ is bounded by $O(\sqrt{k^3\lambda_k})$ for some constant.
This indicates an asymptotic upper bound for the normalized cut of clusters, associated with eigenvalues $\lambda_k$ and $\lambda_{k+1}$ of $\LM(G)$.

Recall that the matrix $\LS$ aggregated by the weighted sum of Laplacian matrices $\LM_i$ over all $r$ views in $\GS$ in \eqref{eq:mvlap} should preserve the community information of all nodes in $\VS$. Suppose that the nodes in $\GS$ are in $k$ clusters. To preserve the well-formed community structures with low normalized cut bounded by Corollary \ref{corollary:eigengap},   the aggregated matrix $\LS$ should have as small $\frac{\lambda_{k}(\LS)}{\lambda_{k+1}(\LS)}$ as possible.

Consequently, our \textit{eigengap objective} is to find desirable view weights $w_i$ to obtain an $\LS$ that \textit{minimizes} the eigengap function $g_k(\LS)$ in \eqref{eq:gap-obj} based on the spectrum of $\LS$.
\begin{equation}\label{eq:gap-obj}
    g_k(\LS) = \frac{\lambda_{k}(\LS)}{\lambda_{k+1}(\LS)}
\end{equation} 

\begin{figure}[!t]
    \captionsetup[subfloat]{labelfont={scriptsize}, textfont={scriptsize}}
    \vspace{-1.5mm}
    \subfloat[{Graph views $G_1$ and $G_2$ with 8 nodes in 2 clusters $\CS_1$ and $\CS_2$.}]{
    \includegraphics[width=0.43\columnwidth]{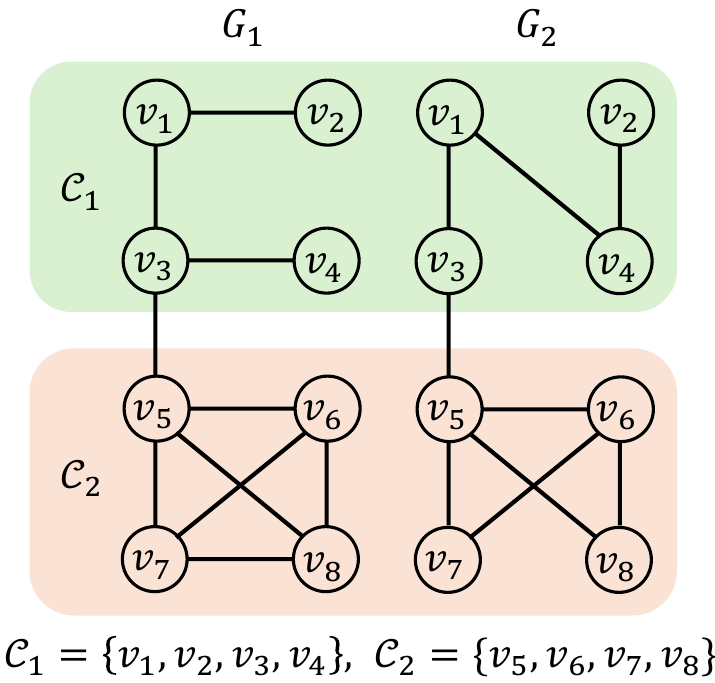}
    \label{fig:obj-example-a}}
    \quad
    \subfloat[Objective values with varied view weights $w_1,w_2$.]{
    \resizebox{0.46\columnwidth}{!}{%
    \vspace{-5mm}
    \begin{tabular}[b]{|cc|c|c|c|} 
    \hline
    \setlength{\tabcolsep}{0.2pt}
      {$w_1$} & {$w_2$} & $g_k(\LS)$ & $\lambda_2(\LS)$ &$g_k- \lambda_2$\\ \hline
      1.0 &0.0 &0.280 &0.148 &0.132 \\ 
      0.9 &0.1 &0.242 &0.158 &0.084 \\
      0.8 &0.2 &0.213 &0.166 &0.047 \\
      0.7 &0.3 &0.191 &0.171 &0.020 \\
      0.6 &0.4 &0.178 &0.174 &0.004 \\
      0.5 &0.5 &0.186 &0.176 &0.010 \\
      0.4 &0.6 &0.202 &0.176 &0.026 \\
      0.3 &0.7 &0.221 &0.174 &0.047 \\
      0.2 &0.8 &0.243 &0.171 &0.072 \\
      0.1 &0.9 &0.269 &0.166 &0.103 \\
      0.0 &1.0 &0.299 &0.159 &0.140 \\\hline
    \end{tabular}
}
    \label{fig:obj-example-b}
    }

    \vspace{-1mm}
    \hspace{-3mm}
    \subfloat[$\LM_1$ \!($w_1=1, w_2=0$).\!]{
    \hspace*{0.3cm} 
    \includegraphics[width=0.23\columnwidth]{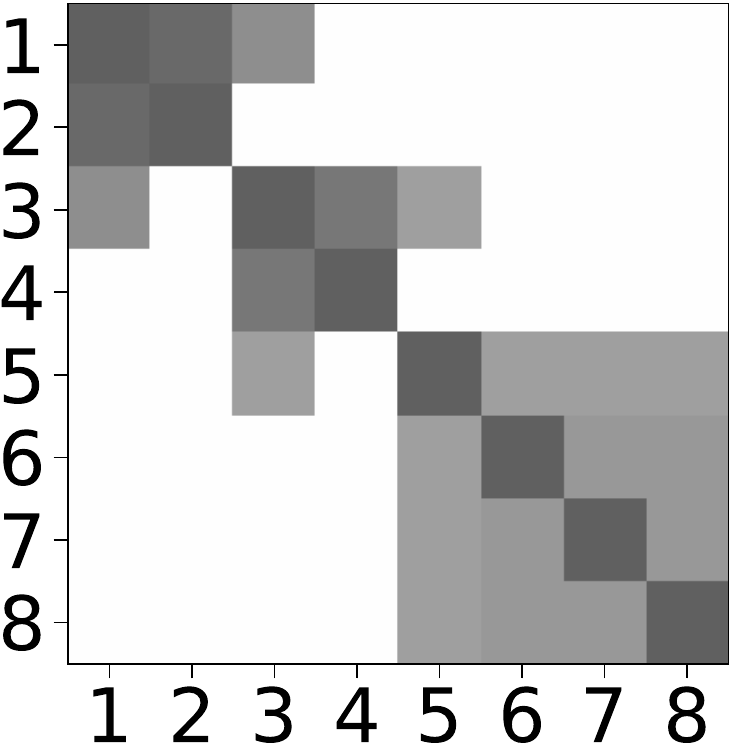}
    \label{fig:lap-1}
    \hspace*{0.3cm} 
    }
    \subfloat[$\LM_2$ \!($w_1=0, w_2=1$).\!]{
    \hspace*{0.2cm} 
    
    \includegraphics[width=0.23\columnwidth]{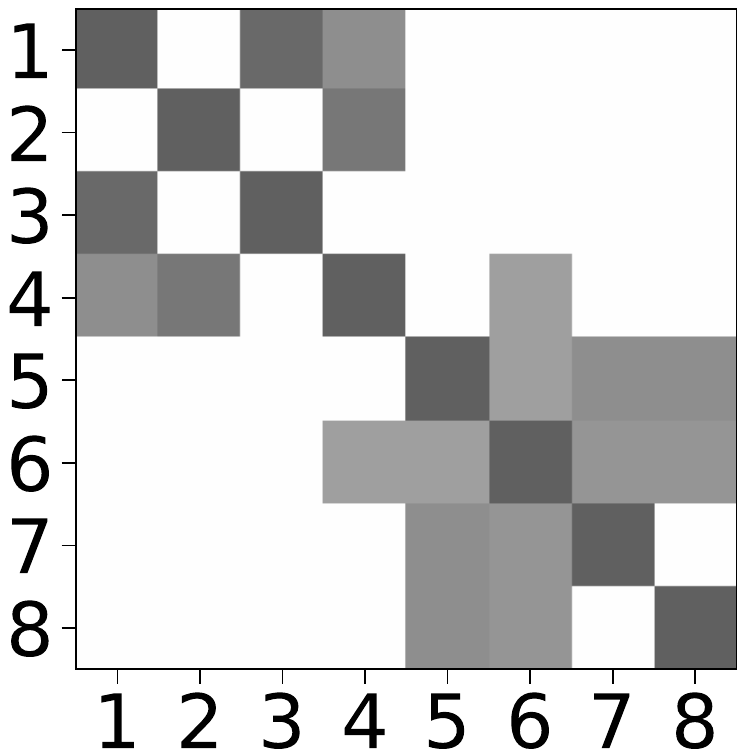}
    \hspace*{0.2cm} 
    \label{fig:lap-2}
    }
    \subfloat[$\LS$ \!($w_1=0.6, w_2=0.4$).\!]{
    \hspace*{0.2cm} 
    \includegraphics[width=0.23\columnwidth]{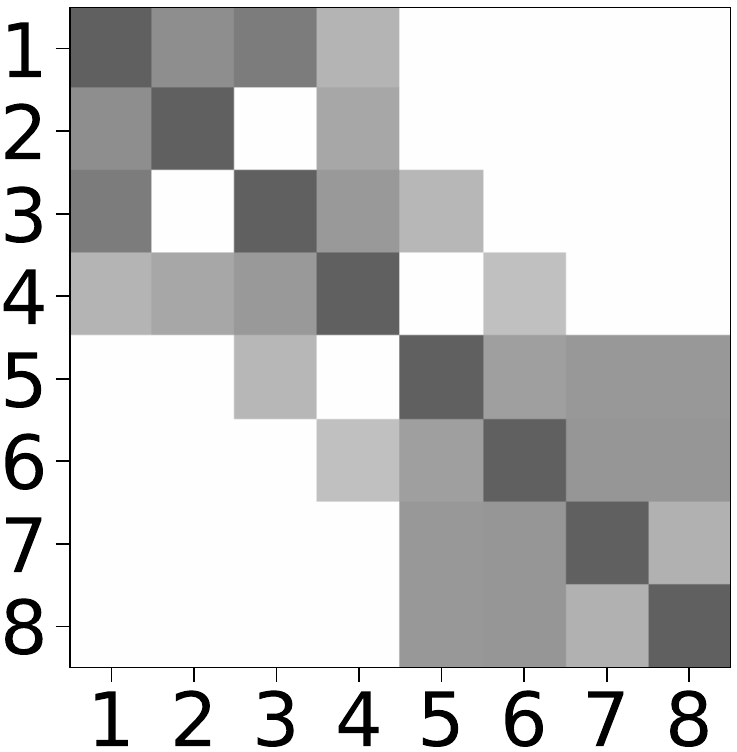}
    \label{fig:lap-L}
    \hspace*{0.3cm} 
    }
    \vspace{-1mm}
    \caption{A running example.}
    \label{fig:obj-example}
    \vspace{-4mm}
\end{figure}

\vspace{0.6mm}
\begin{example} \revision{We use the two graph views $G_1, G_2$ shown in Figure \ref{fig:obj-example-a} as an \mvag example for illustrating the eigengap objective. }
$\GS$ contains 8 nodes in two ground-truth clusters $\CS_1=\{v_1,v_2,v_3,v_4\}$ and $\CS_2=\{v_5,v_6,v_7,v_8\}$,  illustrated by different colors.
Observe that, when only considering a single graph view, either $G_1$ or $G_2$, $\CS_1$ does not exhibit a clear cluster structure due to the sparse connections in it per graph view, while the structure of $\CS_2$ is clear in a single view.
The observation is confirmed by the visualization of Laplacian matrices $\LM_1$ and $\LM_2$ of $G_1$ and $G_2$ in Figure \ref{fig:lap-1}-\ref{fig:lap-2}, where values of larger magnitude are darker: the block formed by nodes $v_1$-$v_4$ in $\CS_1$ are not cohesive, while the values for nodes $v_5$-$v_8$ illustrate a clear cluster for $\CS_2$.
The second column of Table \ref{fig:obj-example-b} shows the eigengap values $g_k(\LS)$ when varying view weights $w_1$ and $w_2$ \revision{for aggregating the two view Laplacians.}
When assigning a large weight to a single view, the eigengap objective of the constructed $\LS$ is large (e.g. 0.242 when $w_1=0.9,w_2=0$ or 0.269 when $w_1=0.1,w_2=0.9$), indicating that the clusters are not well-preserved by $\LS$, due to the observations made above in Figure \ref{fig:obj-example-a}, \ref{fig:lap-1} and \ref{fig:lap-2}.
The eigengap objective is reduced to 0.178 when $w_1=0.6,w_2=0.4$, and we visualize the corresponding $\LS$ in Figure \ref{fig:lap-L}, in which, both clusters $C_1$ and $C_2$ can be clearly observed. Besides, the eigengap obtained by equal weights is 0.186, larger than 0.178. The example shows the intuition for minimizing the eigengap $g_k(\LS)$ to obtain proper view weights.

\end{example}

\subsection{Connectivity Objective}\label{sec:connectivityobj}
The inherent connectivity of graphs is fundamental to the efficacy of graph algorithms~\cite{nagamochi2008algorithmic}.
Nevertheless, some graph views contain connection bottlenecks or leave certain nodes unconnected.
Therefore, the \mvag Laplacian matrix $\LS$ in \eqref{eq:mvlap} should amalgamate the connectivity of all views in $\GS$. To this end, we formulate a connectivity objective that exploits the association between conductance and the spectrum of $\LS$.

Given a simple graph $G$, its conductance $\Phi(G)$ measures how fast a random walk on $G$ converges to its stationary distribution. In \eqref{eq:conductance}, $\Phi(G)$ is defined by the minimum normalized cut of the smaller side in all possible partitions. 
\begin{equation}\label{eq:conductance}
    \Phi(G) = \min_{\CS\mid \vol(\CS)\leq \vol(\VS)/2} \frac{\cut(\CS)}{\vol(\CS)}.
\end{equation}

{Higher graph conductance indicates stronger connectivity. However, computing the exact conductance is intractable in practice. 
From spectral graph theory~\cite{spielmanSpectralGraphTheory2012}, the graph conductance is bounded with $\lambda_2$, the second smallest eigenvalue of the normalized graph Laplacian.}
\begin{equation}
    \frac{\lambda_2}{2} \leq \Phi(G) \leq \sqrt{2\lambda_2}.
\end{equation}

Obviously, a substantial $\lambda_2$ guarantees a lower bound for conductance.
Therefore, we propose the \textit{connectivity objective} to \textit{maximize} the second smallest eigenvalue of $\LS$, denoted by $\lambda_2(\LS)$, to preserve the overall connectivity in $\GS$.

\vspace{0.6mm}
\begin{example}
Recall that, in Figure \ref{fig:obj-example-a},  cluster $\CS_1$ has weak connectivity inside itself in both graph views $G_1$ and $G_2$ of $\GS$. 
The third column in Table \ref{fig:obj-example-b} shows the connectivity objective values $\lambda_2(\LS)$ of the obtained $\LS$ when varying view weights. 
When a single view has a large weight,  $\lambda_2(\LS)$  is smaller, indicating weak connectivity, which matches the observation above. 
If $\lambda_2(\LS)$ is sufficiently large, e.g., 0.174 when $w_1=0.6,w_2=0.4$, the obtained $\LS$ can preserve the connectivity information from both views in $\GS$, as illustrated in the corresponding visualization in Figure \ref{fig:lap-L}, where the values for $v_1,v_2,v_3,v_4$ are clear to represent cluster $\CS_1$.

\end{example}

\subsection{The Full Objective} \label{sec:overall-obj}
In \eqref{eq:fullobj}, we combine the eigengap objective $g_k(\LS)$ and connectivity objective $\lambda_2(\LS)$ into the full objective function $\fun(w_1,...,w_r)$, also denoted by $h(\wv)$ where weight vector $\wv=[w_1,\dots, w_r]$.
Since the view weights should minimize $g_k(\LS)$ while maximizing $\lambda_2(\LS)$,  in  \eqref{eq:fullobj}, $\lambda_2(\LS)$ takes a negative sign, so that $\fun(\wv)$ is to be minimized. 
To prevent $\LS$ from being dominated by a single view, we introduce a regularization term of all weights with parameter $\gamma$. 
\begin{equation}\label{eq:fullobj}
 \fun(\wv)=\fun(w_1,\dots,w_\nv) = g_k(\LS) - \lambda_2(\LS) + \gamma \sum\limits_{i=1}^{\nv} w_i^2
\end{equation}

The objective function $h$ estimates the suitability of $\LS$ for performing MVAG analytics and guides the search for appropriate view weights. Therefore, our problem statement in Section \ref{sec:problem} is formulated by a constrained optimization problem, aiming to find the optimal weights $\wv^*\in \mathbb{R}^r$ that minimize $h(\wv)$ while satisfying the constraints.
\begin{equation}\label{eq:constraints}
\begin{aligned}
    \wv^*&=\argmin_{\wv} \fun(\wv)=\argmin_{w_1,\dots, w_{\nv}} \fun(w_1,\dots, w_{\nv}), \\ 
    \textrm{s.t. } &  \textrm{any } w_i\geq 0\textrm{ and } \sum\limits_{i=1}^{r}w_i = 1 \textrm{.}
\end{aligned}
\end{equation}

\stitle{Discussion}
Although both eigengap and connectivity objectives have associations with the measure of normalized cut, they each focus on different aspects of \mvags.
The eigengap objective is linked with the existence of multiple well-formed clusters, while the connectivity objective prevents nodes from being isolated from the main graph structure.
\revision{Our combination of the two objectives aim to preserve both properties, achieving a balance between them.
In the experiments, our method with the full objective in \eqref{eq:constraints}, combining both objectives,  consistently outperforms approaches using a single objective.}

{\begin{figure}[!t]

\centering
\definecolor{darkred}{RGB}{139,0,0}

\begin{tikzpicture}
    \begin{customlegend}[
    legend entries={sampled point, optimal weights},
    legend columns=2,
    legend style={at={(0.5,1.05)},anchor=north,draw=none,font=\small,column sep=0.2cm}]
    \addlegendimage{only marks, color=black, mark=*}
    \addlegendimage{only marks, color=black, mark=x}
\end{customlegend}
\end{tikzpicture}
\vspace{-4.6mm}
\\[-\lineskip]

\resizebox{0.44\columnwidth}{!}{%
\subfloat[Objective function $\fun$]{
\label{fig:add-yelp}
\resizebox{0.44\columnwidth}{!}{%
\includegraphics{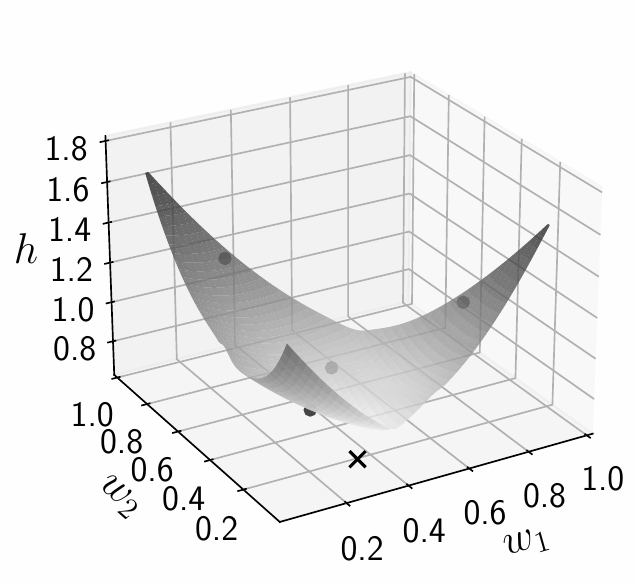}}
}
}
\hspace{0.6em}
\resizebox{0.44\columnwidth}{!}{%
\subfloat[Quadratic interpolation $\qopt$]{
\label{fig:linear-yelp}
\resizebox{0.44\columnwidth}{!}{%
\includegraphics{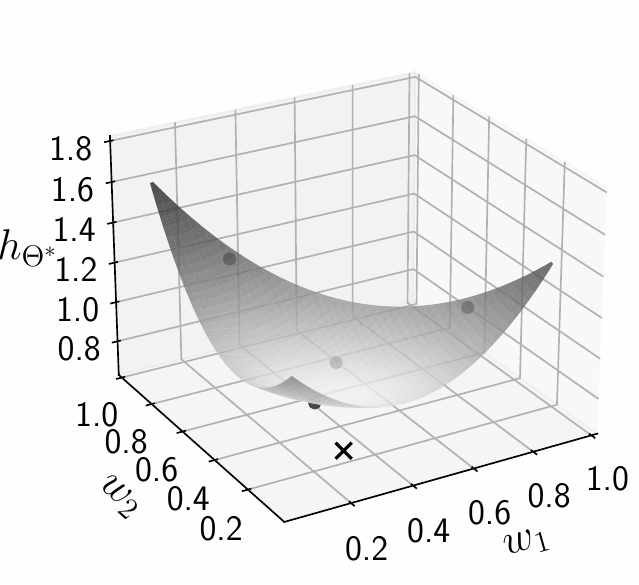}}
}
}

\vspace{-1mm}
\caption{{Plot of objective functions on Yelp.}} \label{fig:obj-plots}
\vspace{-4mm}
\end{figure}
}

\vspace{0.6mm}
\begin{example}\label{example:fullobj} 
{
To visualize the distribution of the objective $\obj(\wv)$  over all possible view weights, a case study is performed on Yelp dataset with three views (see Table \ref{tab:datasets} for details). Among the three view weights $w_1,w_2,w_3$, we vary $w_1$ and $w_2$ at interval $0.01$ and set $w_3 = 1-w_1-w_2$, to exhaust all possible weight combinations, and we plot the value of $\obj(w_1,w_2,w_3)$ in Figure \ref{fig:add-yelp}. The plot shows a generally smooth surface that curves downward, which visually demonstrates the suitability of the proposed formulation. 
}

\end{example}

\section{Algorithms}\label{sec:algorithms}
In Section \ref{sec:basemethod}, we develop the \ours method, which optimizes the objective to determine view weights. \ours has demonstrated superior  performance compared to existing methods. To further enhance efficiency while preserving result quality, we present \qours in Section \ref{sec:pi}.

\subsection{\ours Method}\label{sec:basemethod}
\revision{Given an \mvag with $r$ view Laplacians, the search space for possible view weights is exponential, rendering an exhaustive grid search intractable for solving \eqref{eq:constraints}.
Optimizing the non-convex objective function $\fun(\wv)$ is further complicated by the presence of both inequality and equality constraints.
Additionally, evaluating $h(\wv)$ and its gradients is costly due to intensive eigenvalue computations. Traditional gradient-descent methods are inefficient, as they require many iterations to converge and incur significant computational overhead.}

\begin{figure}[!t]

\centering

\includegraphics[width=0.99\columnwidth]{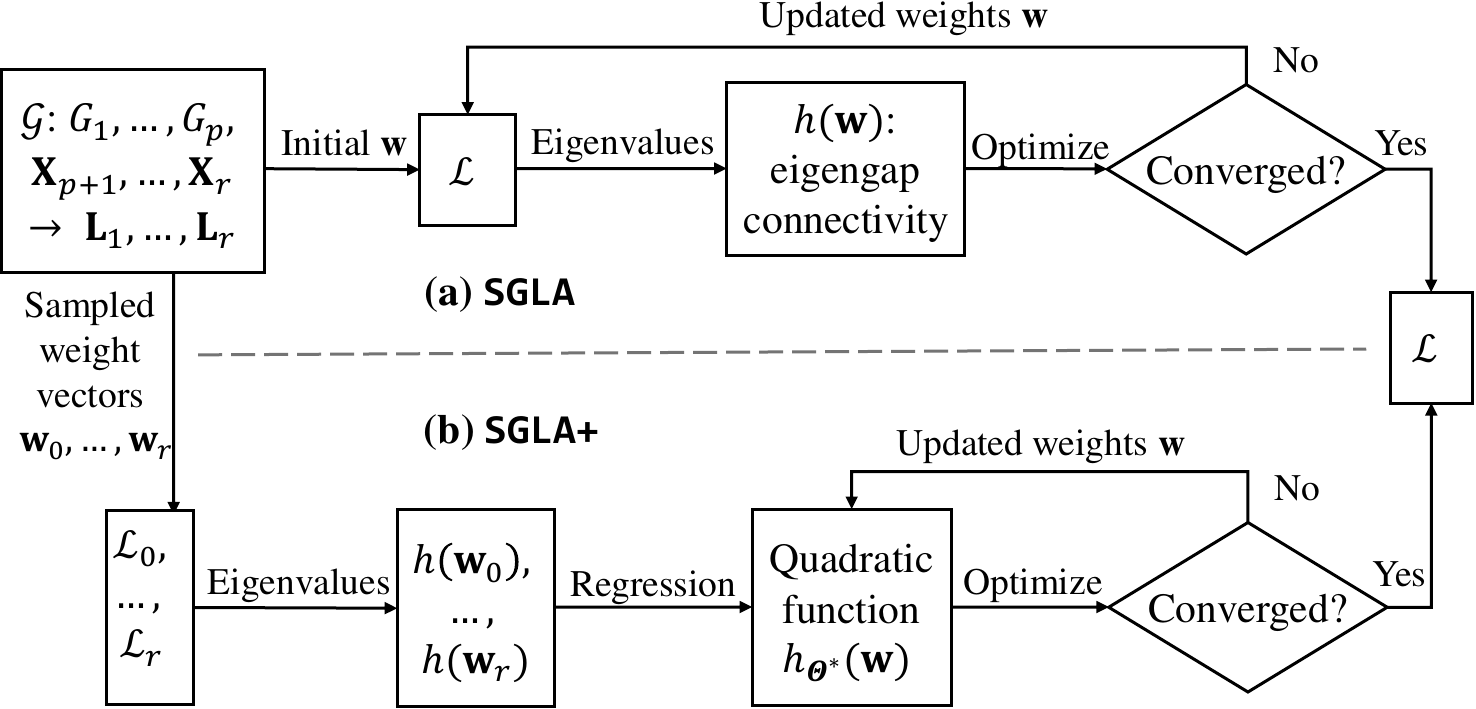}
\caption{Overview of \main and \qours algorithms.} \label{fig:flow}
\vspace{-0mm}
\end{figure}

\revision{To address the technical challenges, we develop the base method \ours that produce high-quality results in a reasonable amount of time.}
\revision{\ours iteratively performs two key computations: (i) \textit{objective evaluation} and (ii) \textit{objective optimization to update view weights}.} 
Figure \ref{fig:flow}a provides an illustration of \ours. 
After obtaining the Laplacian matrices $\LM_1,...,\LM_r$ of the $r$ graph views and attribute views in the input $\GS$ as explained in Section \ref{sec:problem} and initializing $\wv$ with uniform weights,  
\revision{During objective evaluation,} \ours computes the latest $\LS$ and the eigenvalues of $\LS$ first and evaluates the objective function $h(\wv)$. \revision{Then, to update weights,} \ours optimizes and updates $\wv$ via a derivative-free optimizer.
This optimizer guarantees a local optimum when the variables in $\wv$ converge.
\ours terminates either when the update of view weights is negligible, \ie convergence, or when the number of iterations exceeds a limit. \ours eventually returns $\LS$, which will be used for clustering and embedding tasks. 

\vspace{0.6mm}
\noindent\textbf{Algorithm.}
The pseudo code of \ours is displayed in Algorithm \ref{alg:direct}. 
The weight parameters $w_1,...,w_r$ of all $r$ views are initialized to $\frac{1}{r}$ at Line 1. 
\revision{Then, from Lines 2 to 9, \ours performs objective evaluation (Lines 3-5) and optimizes the proposed objective to update weights (Lines 6-9)}, in an iterative fashion for at most $T_{max}$ iterations, and early terminates if the  condition at Line 7 is met.
Specifically, in an iteration, with the current weight parameters $w_i$, we first obtain the aggregated matrix $\LS$ by weighted sum of the Laplacian matrices of all $r$ views at Line 3. 
Then at Line 4, we compute the $k+1$ smallest eigenvalues of $\LS$.
Note that all matrices, including $\LM_i$ and $\LS$, are organized as sparse matrices, and thus the computation at Lines 3 and 4 can be efficiently processed.
With the eigenvalues $\lambda_2$, $\lambda_k$, and $\lambda_{k+1}$, we can compute the eigengap and connectivity objectives formulated in Section \ref{sec:sgfobjective} to obtain $\obj(w_1,\dots,w_{\nv})$ by \eqref{eq:fullobj} (Line 5).
$\Omega$ at Line 6 represents the constraints specified in \eqref{eq:constraints}.
At Line 6, we adopt the %
optimizer  \cobyla~\cite{powellDirectSearchOptimization1994}, considering the objective function and constraints, to update the first $r-1$ view weights to $w'_1,\dots,w'_{r-1}$ as $w'_r$ follows trivially. 
Briefly, this optimizer performs interpolation in the direction of each variable and updates a regular-shaped simplex of variables over iterations,  as the trust region that conforms to the constraints while minimizing the objective. The process stops when the updated variables are close to the previous iteration. 
At Line 7, if the difference between the updated view weights $w'_1,\dots,w'_{r-1}$ and the previous weight values $w_1,\dots,w_{r-1}$ measured by Euclidean distance, is below an early termination criteria $\epsilon$,  the iterative process of \ours terminates and returns. Otherwise, we need to update the latest weights to $w_1,\dots,w_{r-1}$ and $w_r$ (Lines 8-9), and continue with the next iteration.
Finally, the \mvag Laplacian matrix $\LS$ is returned at Line 10, to be used in downstream tasks, including embedding and clustering.

\begin{algorithm}[!t]
\caption{\ours} \label{alg:direct}
\small
\KwIn{View Laplacians $\LM_1, \dots, \LM_\nv$ of the input \mvag $\GS$, number of clusters $k$, constraints $\cons$, algorithm parameters $\tmax, \tol$.}
 Initialize weight parameters $w_1,\dots,w_{\nv}\gets \frac{1}{\nv}$ \;
 \For{$t\gets 1, \dots, \tmax$}{
 $\LS \gets \sum\limits^\nv_{i=1} w_i \LM_i$\;
 $\lambda_1,\dots,\lambda_{k+1}\gets \text{Eigenvalues}(\LS, k+1)$\;
 
 Obtain $\obj(w_1,\dots,w_{\nv})$ by \eqref{eq:fullobj}\;
 {$w_1^\prime,\dots,w_{\nv-1}^\prime\gets \cobyla(\obj(w_1,\dots,w_{\nv}), \cons)$\;}
 \lIf{$\sqrt{\sum\limits_{i=1}^{\nv-1} (w_i^\prime-w_i)^2} <\tol$}{\textbf{break}}
 $w_1, \dots, w_{\nv-1}\gets w_1^\prime,\dots,w_{\nv-1}^\prime$\;
 $w_\nv\gets 1- \sum\limits^{\nv-1}_{i=1}w_i$\;
 }
 \Return $\LS$\;
\end{algorithm}

\vspace{0.6mm}
\noindent\textbf{Complexity.} Given an \mvag $\GS$ with $n$ nodes and $r$ views,  
all view Laplacian matrices are stored in sparse matrix format, and thus the costs of additions and matrix-vector multiplication operations on $\LS$ are linear to the count of nonzero elements, \ie at most $2(m+qnK)$.
The aggregation in Line 3 includes scalar multiplications and additions on Laplacian matrices. Line 4 solves eigenvalues via a bounded number of matrix-vector multiplications.
Therefore, Lines 3-4 incur $O(m+qnK)$ time combined.
Lines 5-6 and 8 are simple $O(\nv)$ computations, while the optimizer in Line 7 takes $O(\nv^2)$ time to update $\nv-1$ variables. 
$r$ is typically small and can be regarded as a constant. 
When $T$ iterations are conducted, we conclude that the overall time and space complexity of \ours is $O(T(m +qnK))$.

\subsection{\qours Method}\label{sec:pi}
\revision{Although \ours outperforms existing methods, it encounters significant computational challenges on large-scale \mvags.
The main bottleneck is the costly evaluation of the objective function $\fun(w_1,\dots,w_\nv)$ (Lines 3-5 in Algorithm \ref{alg:direct}), which involves intensive eigenvalue computations. This evaluation must be repeated across many iterations for \ours to converge, resulting in substantial overhead that limits its scalability.}

To further improve efficiency,  we design \qours that has lower complexity than \ours. 
\revision{\qours has the following key designs. (i) Instead of directly optimizing the objective $h(\wv)$,  \qours formulates and optimizes an approximation $\qreg$ of $h(\wv)$. This approximation is constructed as a quadratic interpolation that is quick to evaluate while closely resembling the original $h(\wv)$. (ii) We develop a sampling strategy to obtain $(r+1)$  weight vectors
as samples of $h(\wv)$ to find an accurate approximation $\qopt$, requiring only  ($r+1$) objective evaluations,  fewer than those required by \ours.
(iii) Finally, \qours efficiently minimizes   $\qopt$ to compute the desired view weights necessary for constructing the MVAG Laplacian  $\LS$. Figure \ref{fig:flow}b provides an  overview of \qours. Below, we first formulate the approximation of our objective, explain the sampling strategy, and then present the algorithm details.}

\vspace{0.6mm}
\noindent
\revision{\textbf{Objective Approximation.}}
The objective $\fun(w_1, \dots, w_{\nv})$  in \eqref{eq:constraints} has a weight vector $\wv\in\mathbb{R}^r$ with  $r$ elements $w_i$, for $1\leq i\leq r$, as variables. 
On the other hand, in optimization with a univariate objective, it is possible to find an approximate minimum by fitting a quadratic polynomial to three values of the objective~\cite{heathScientificComputingIntroductory2018}. 
Thus, we generalize quadratic interpolation to multiple variables. 
Specifically, given a set of weight vector samples, we aim to find a function $\qreg(w_1, \dots, w_{\nv})$ with a coefficient set $\ThM$, as the interpolation of $\fun(w_1, \dots, w_{\nv})$.
As shown in \eqref{eq:qreg}, $\qreg$ comprises all second-degree terms of $w_i$ and $w_j$ with coefficients $\theta_{i,j}$, where $1\leq i\leq j\leq\nv-1$, linear terms $w_i$ with coefficients $\theta_{i,r}$ for $i=1,...,r-1$, and a constant term $\theta_{r,r}$. We leave out $w_r$ because it can be determined by the equality constraint that all weights sum up to 1.
The matrix format of $\qreg$ is in \eqref{eq:qregMat}, where the upper triangular matrix $\ThM$ consists of non-zero entries $\ThM[i,j]=\theta_{i,j}$ for $i,j\in{1,...,r}$ and $i\leq j$. 
\begin{align}
  \qreg(\wv)&= 
    \sum\limits_{1\leq i\leq j\leq\nv-1}\theta_{i, j} w_i w_j + \sum\limits_{i=1}^{\nv-1}\theta_{i, \nv} w_i + \theta_{\nv,\nv} \label{eq:qreg}\\
    \qreg(\wv)& = \rvect{w_1, \dots, w_{\nv-1}, 1}\transpose \ThM \rvect{w_1, \dots, w_{\nv-1}, 1} \label{eq:qregMat}
\end{align}

To determine the coefficients in $\ThM$, we perform regression with $(r+1)$ weight vector samples $\wv_\ell$, for $0\leq \ell\leq r$. Sampling details will be explained shortly.
Our goal  is to find a solution $\ThM^*$, such that the squared error between $\fun(\wv_\ell)$ and $\qreg(\wv_\ell)$ over the $(r+1)$  samples is minimized in \eqref{eq:ridge}. We address this problem with a least Frobenius norm quadratic model~\cite{ragonneauOptimalInterpolationSet2024}.
Specifically, the optimization function in \eqref{eq:ridge} exerts regularization $\|\ThM\|^2_F$ with parameter $\ralpha$, to minimize the Frobenius norm of the coefficient matrix $\ThM$, in order to find the desired solution. 
\begin{equation}\label{eq:ridge}
    \ThM^*=\argmin_{\ThM} \left(\sum\limits_{\ell=0}^{\nv}\big(\fun(\wv_\ell)-\qreg(\wv_\ell)\big)^2 + \ralpha\|\ThM\|^2_F %
    \right)
\end{equation}

After evaluating the objective function $\fun(\wv_\ell)$ for each $\wv_\ell$ among the $(r+1)$ weight vector samples, the regression in \eqref{eq:ridge} can be solved via Cholesky decomposition. With coefficients $\ThM^*$ found, the function $\qopt$ is a local approximation of $\fun$. Subsequently, 
we find the minimum solution $\wv^\dagger$ of $\qopt$ under constraints, \ie \eqref{eq:regcons}, and $\wv^\dagger$ is regarded as an approximate solution for the original problem in \eqref{eq:constraints}.
This procedure is much more efficient than direct optimization of $h(\wv)$ in \ours, as the evaluation of $\qopt$ does not require the construction of $\LS$ or the computation of eigenvalues. 
\begin{equation} \label{eq:regcons}
    \small
    \wv^\dagger=\argmin_{\wv}   \qopt(\wv) 
     \textrm{ s.t. }    w_i\geq 0\;\forall\;1\leq i\leq r\textrm{, } \sum\limits_{i=1}^{r}w_i = 1 \textrm{.}
\end{equation}

\vspace{0.6mm}
\noindent
\revision{\textbf{Weight Vector Sampling.}}
The coefficient matrix $\ThM$ contains $\frac{\nv(\nv+1)}{2}$ nonzero entries. 
Reaching a unique solution of $\ThM$ requires $O(\nv^2)$ weight vector samples for expensive objective evaluations. Instead, we propose a scheme to gather $(r+1)$ weight vector samples only, which are sufficient in practice to find a high-quality $\qopt$ via \eqref{eq:ridge} and consequently $\wv^\dagger$ in \eqref{eq:regcons}, as validated by experiments. 
The first sample $\wv_0=\rvect{\frac{1}{\nv},\frac{1}{\nv},\dots, \frac{1}{\nv}}\in \RN^{\nv}$ assigns the same weight $\frac{1}{\nv}$ for all $\nv$ views of $\GS$.
Then for each $\ell$-th view, a weight vector $\wv_\ell$ is sampled as the midpoint between $\wv_0$ and the one-hot vector $\mathbf{1}_\ell\in \{0,1\}^r$ that assigns a full weight to the $\ell$-th view, where $1\leq \ell\leq r$.
Specifically, $\wv_\ell=(\wv_0+\mathbf{1}_\ell)/2$, \ie the $\ell$-th element in vector $\wv_\ell$ has value $\frac{r+1}{2r}$ while the other elements have value $\frac{1}{2r}$.

  \begin{algorithm}[!t]
\caption{\qours} \label{alg:1} 
\small
\KwIn{View Laplacians $\LM_1, \dots, \LM_\nv$ of the input \mvag $\GS$, number of clusters $k$, constraints $\cons$, algorithm parameters $\ralpha, \tmax, \tol$.}
 $\wv_0 \gets \rvect{\frac{1}{\nv},\frac{1}{\nv},\dots, \frac{1}{\nv}}\in \RN^{\nv}$\;
 \For{$\ell\gets 0, \dots, \nv$}{
  \lIf{$\ell>0$}{\textbf{$\wv_\ell\gets (\wv_0+\mathbf{1}_\ell)/2$}}
 Obtain $\LS_\ell$ by $\wv_\ell$ and $\LM_i$ via \eqref{eq:mvlap}\; 
 $\lambda_1,\dots,\lambda_{k+1}\gets \text{Eigenvalues}(\LS_\ell, k+1)$\;
 Obtain $\fun{(\wv_\ell)}$ by \eqref{eq:fullobj}\;
 }
 Solve $\ThM^*$ in \eqref{eq:ridge} for observations $(\wv_0, \fun{(\wv_0)}),\dots, (\wv_\nv, \fun{(\wv_\nv)})$ and L2 multiplier $\ralpha$\;
 $w_1,\dots,w_{\nv}\gets \frac{1}{\nv}$ \;
 \For{$t\gets 1, \dots, \tmax$}{
 Calculate $\qopt(w_1,\dots,w_{\nv})$ by \eqref{eq:qregMat}\;
 {$w_1^\prime,\dots,w_{\nv-1}^\prime\gets \cobyla(\qopt(w_1,\dots,w_{\nv}), \cons)$, where $\cons$ represents the constraints in \eqref{eq:regcons}\;}
 \lIf{$\sqrt{\sum\limits_{i=1}^{\nv-1} (w_i^\prime-w_i)^2} <\tol$}{\textbf{break}}
 $w_1, \dots, w_{\nv-1}\gets w_1^\prime,\dots,w_{\nv-1}^\prime$\;
 $w_\nv\gets1- \sum\limits^{\nv-1}_{i=1}w_i$\;
 }
 $\LS \gets \sum\limits^\nv_{i=1} w_i \LM_i$\;
 \Return $\LS$\;
\end{algorithm}

\vspace{0.9mm}
\begin{example} {
For the Yelp dataset containing three views ($r=3$), the objective function $\fun$ is plotted in Figure \ref{fig:add-yelp} in Example \ref{example:fullobj}. 
Following the sampling scheme explained above, we first obtain 4 weight vector samples $\wv_0=\rvect{\frac{1}{3}, \frac{1}{3},\frac{1}{3}}$, $\wv_1=\rvect{\frac{2}{3}, \frac{1}{6}, \frac{1}{6}}$, $\wv_2=\rvect{\frac{1}{6}, \frac{2}{3}, \frac{1}{6}}$ and $\wv_3=\rvect{\frac{1}{6}, \frac{1}{6}, \frac{2}{3}}$. Each of $\wv_1, \wv_2, \wv_3$ emphasizes a specific view in the Yelp dataset.
Figure \ref{fig:add-yelp} marks the locations of these four sampled points w.r.t. the weights of the first two views, as the third view weight is determined by the equality constraint that all weights add up to 1.
Observe that the plot of $\fun$ in Figure \ref{fig:add-yelp} resembles a partial paraboloid surface. According to \eqref{eq:qregMat}, we can get $\qreg$ %
on the Yelp dataset.
The coefficients $\ThM^*$ are determined by solving \eqref{eq:ridge}. We plot the acquired interpolation $\qopt$ in Figure \ref{fig:linear-yelp}, which exhibits a paraboloid surface similar to the original objective $\fun$. In addition, we use crosses in Figure \ref{fig:add-yelp} and \ref{fig:linear-yelp} to mark the weights that minimize $\fun$ and $\qopt$, respectively. Their close locations validate that $\qopt$ is an effective approximation for minimizing $\fun$.
}
\end{example}

\stitle{Algorithm}
{The pseudo code of \qours is provided in Algorithm \ref{alg:1}. 
\revision{From Lines 1 to 7, we sample ($r+1$) weight vectors,  evaluate the objective $\obj(\wv_\ell)$ over the sampled weight vectors, and solve $\ThM^*$ to get the objective approximation  $\obj_{\ThM^*}$ at Line 7.}
\revision{From Lines 8 to 14, we aim to optimize $\obj_{\ThM^*}$} by  iteratively updating the view weights $w_1,\dots,w_r$, and then the view weights are used to obtain $\LS$ at Line 15.
Specifically, the sampled weight vectors are first calculated, including the equal weights in $\wv_0$ (Line 1), and then each $\wv_\ell$ for $1\leq \ell\leq r$ (Line 2-3). From each sampled vector $\wv_\ell$, we construct the corresponding \mvag Laplacian $\LS_\ell$ at Line 4. After solving its bottom $k+1$ eigenvalues (Line 5), the objective $\obj(\wv_\ell)$ is evaluated at Line 6 as the observation for sample $\wv_\ell$.
At Line 7, interpolation over these $\nv+1$ samples is solved by regression with parameter $\ralpha$ in \eqref{eq:ridge} to obtain $\ThM^*$.
After initializing the target view weights $w_1,...,w_r$ at Line 8, from Lines 9 to 14, we iteratively optimize the acquired interpolation $\qopt$ to update $w_1,...,w_r$.
At each iteration (Line 9),  $\qopt(w_1,...,w_r)$ is efficiently calculated at Line 10, without expensive construction of $\LS$ or eigenvalue computation. 
Updating weights (Lines 11 to 14) and convergence condition (Line 12) are similar to \ours, but are for the optimization of \eqref{eq:regcons} instead of \eqref{eq:constraints}. 
After the termination of iterations, the converged weight parameters are used to construct $\LS$ at Line 15.}

\vspace{0.6mm}
\noindent
\textbf{Complexity.}
At Lines 1-6, \qours performs exactly ($r+1$) objective evaluations over all sampled weight vectors. The time and space complexity of this part is $O(\nv(m+qnK))$.
To solve the $O(\nv^2)$ coefficients in $\ThM$, Line 7 conducts a decomposition that theoretically takes $O(\nv^6)$ time. Since the number of views $r$ is usually less than 10 and regarded as a constant, the cost of Line 7 can be considered negligible in practice.   
In Lines 9-14, Line 10 incurs $O(\nv^2)$ cost for evaluating $\qopt$, and Lines 11-13 also have an $O(\nv^2)$ complexity combined.
Hence, solving the weights over $T$ iterations incurs negligible $O(Tr^2)$ cost.
With $\nv$ as constant, \qours has an overall time and space complexity $O(m+qnK)$, as the processing on graph Laplacians is removed from the optimization loop. Observe that the complexity $O(m+qnK)$ of \qours improves over the complexity $O(T(m +qnK))$ of \ours. Empirically, we also find \qours faster than \ours in experiments.

\section{Experiments}
\label{sec:experiments}

We evaluate the effectiveness and efficiency of our \ours and \qours over 12 competitors for clustering and 8 competitors for embedding on 8 real-world \mvag datasets. Section \ref{sec:expsetup} describes experimental settings. Section \ref{sec:expcluster} and \ref{sec:expembedding} report the results of clustering and embedding, respectively. Section \ref{sec:ablation-study} conducts experimental analysis.

\subsection{Experimental setup}\label{sec:expsetup}

\begin{table}[!t]
  \caption{Statistics of multi-view attributed graph datasets.}
  \label{tab:datasets}
  \vspace{-1mm}
  \centering
    \resizebox{1\columnwidth}{!}{%
    \setlength{\tabcolsep}{3pt}
  \begin{tabular}{llllll}
    \toprule
    Dataset     & $n$  &$\nv$ & 
     $m_i$ of $G_i$  
    & $d_j$ of  $\XM_{j}$     & $k$ \\
    \midrule
    
    {RM} & {91} &11 & \begin{tabular}{@{}l@{}}267; 404; 298; 317; 163; 1,595; \\ 1,683; 1,910; 1,565; 1,044\end{tabular}
       & {32}  & {2}\\

    {Yelp} & {2,614} &3 & 262,859;  1,237,554  &82 &3\\

   {IMDB} & {3,550} &3 & 5,119; 31,439 & {2,000}  & {3}\\
     {DBLP} & {4,057} &4 & 3,528; 2,498,219; 3,386,139 & {334}   & {4} \\

    {{Amazon photos}}  & {7,487} &3 &  119,043 & {745; 7,487}  & {8}\\
    {{Amazon computers}} & {13,381} &3 &  245,778 & {767; 13,381}  & {10}\\
    {MAG-eng} & {1,798,717} &4 &  43,519,012;  10,112,848 & {1,000; 1,000}  & {55}\\
    {MAG-phy} & {2,353,996} &4 &  257,706,767; 18,055,930 & {1,000; 1,000}  & {22}\\

    \bottomrule
  \end{tabular}
  }
  \vspace{-4mm}
\end{table}

\noindent
\textbf{Datasets.} Table \ref{tab:datasets} provides the statistics of  the 8 multi-view attributed graph datasets, including the number of nodes ($n$) and views ($r$), the number of edges $m_i$ for each $i$-th graph view (separated by semicolons),  dimension $d_j$ for each $j$-th attribute view (separated by semicolons), and the number of ground truth node classes $k$ that is also considered the number of clusters.
These datasets are \revision{real-world \mvags from diverse domains}, including social activities (RM~\cite{behrouzFirmTrussCommunitySearch2022}), business (Yelp~\cite{luRelationStructureAwareHeterogeneous2019}, Amazon photos and Amazon computers~\cite{panMultiviewContrastiveGraph2021}), movies (IMDB~\cite{parkUnsupervisedAttributedMultiplex2020}), and academic collaborations (DBLP~\cite{fanOne2MultiGraphAutoencoder2020}, MAG-eng and MAG-phy~\cite{bojchevskiScalingGraphNeural2020b}). \revision{
The MAG-eng and MAG-phy datasets exemplify the complexity and the large scale of academic collaboration networks. MAG-phy includes over 2.35 million nodes with four views: two graph views with over 257.7 million and 18.05 million edges, and two high-dimensional attribute views with 1000 dimensions each. These datasets highlight the sparsity of graph views and the high dimensionality of attribute views, typical in real-world applications.
Also, some datasets feature dense graph views, \eg DBLP, while others are sparse, \eg IMDB.
The diversity, sparsity, and edge distributions of the datasets provide a representative testbed for comprehensive evaluation.}
The ground-truth clusters and class labels are obtained from the original data, \eg movie genres in IMDB and product categories in Amazon. In MAG-phy and MAG-eng, nodes are labeled by the subject domain of their publication venues.

\begin{table*}[ht]
\centering
\renewcommand{\arraystretch}{1.02}
\caption{Clustering quality (the top 3 are in blue; darker shades indicate better results).}\vspace{-2mm}
\begin{small}
\resizebox{1\textwidth}{!}{
\setlength{\tabcolsep}{2.3pt}
\begin{tabular}{|=c|+c+c+c+c+c|+c+c+c+c+c|+c+c+c+c+c|+c+c+c+c+c|+c|}
\hline
\multirow{2}{*}{\bf{Method} } & \multicolumn{5}{c|}{\bf{RM}} & \multicolumn{5}{c|}{\bf{Yelp}} & \multicolumn{5}{c|}{\bf{IMDB}} & \multicolumn{5}{c|}{\bf{DBLP}} &\multirow{2}{*}{\shortstack{\bf{Overall}\\ \bf{rank}}}\\ \cline{2-21}
&Acc  &F1 & NMI  & ARI &\revision{Purity} &Acc &F1 & NMI  & ARI &\revision{Purity} &Acc &F1 &NMI  & ARI  &\revision{Purity} &Acc  &F1 & NMI  & ARI &\revision{Purity} & \\ \hline
\wmsc   &0.626	&0.474	&0.001	&-0.017 &\revision{0.703} 
        &0.813	&0.836	&0.537	&0.489 &\revision{0.813} 
        &0.374	&0.291	&0.003	&0.001 &\revision{0.379} 
        &0.780	&0.778	&0.468	&0.507 &\revision{0.780} &9.7\\
\twocmv &0.904	& 0.899 &0.703	 &0.665	 &\revision{0.914}
        &0.857  &0.742  &0.576   &0.592  &\revision{0.857}
        &0.510  &0.486  &0.127   &0.148  &\revision{0.511}
        &0.914  &0.850  &0.749   &0.797	 &\revision{0.914} &6.8\\ 
\mega   &0.802  &0.793  &0.423   &0.359	    
        &\revision{0.802} 
        &0.653	&0.568  &0.390	&0.427	&\revision{0.733} 
        &0.390  &0.239 &0.007   &0.004  &\revision{0.392}
        &0.913	&0.907 &0.741	&0.792	&\revision{0.913} &\revision{8.3}\\ 
\hdmi   &0.613  &0.459 &0.010 &-0.018  &\revision{0.703}
        &0.909  &0.915 &0.681 &0.727  &\revision{0.909}
        &0.541  &0.547 &0.162 &0.142 &\revision{0.532} 
        &0.895  &0.885 &0.706 &0.761 &\revision{0.896} &8.5\\ 
\uramn  &0.736  &0.684 &0.107 &0.195   &\revision{0.736}
        &0.771  &0.762 &0.490 &0.483 &\revision{0.771}
        &\first{0.588} &\first{0.582} &0.183 &0.197 &\first{\revision{0.588}}
        &0.908  &0.901 &0.715 &0.781 &\revision{0.896} &\revision{6.0}\\
\omac   &0.659  &0.397 &0.040 &-0.044 &\revision{0.703}
        &0.649  &0.565 &0.391 &0.425 &\revision{0.732}
        &0.547  &\third{0.550} &0.135 &0.139 &\revision{0.535}
        &0.873  &0.865 &0.669 &0.705 &\revision{0.877} &\revision{9.5}\\ 
\dmg    &0.745  &0.623 &0.147 &0.191 &\revision{0.765}
        &0.714  &0.725 &0.441 &0.365 &\revision{0.714}
        &0.545  &0.459 &\third{0.195} &\third{0.209} &\revision{0.550}
        &0.925  &0.921 &0.761 &0.815 &\revision{0.925} &6.3\\ 
\lmgec  &0.703  &0.500 &0.015 &0.044 &\revision{0.703}
        &\third{0.923} &\third{0.928} &\third{0.725} &\third{0.764} &\third{\revision{0.923}}
        &\second{0.568} &\second{0.577} &0.166 &0.143 &\revision{\second{0.562}}
        &0.922 &0.917  &0.757 &0.813 &\revision{0.922} &\revision{5.1}\\  
\magcn  &0.703 &0.736 &0.000 &0.000 &\revision{0.703}
        &0.734 &0.705 &0.437 &0.455 &\revision{0.734}
        &0.513 &0.482 &0.116 &0.135 &\revision{0.511}
        &- &- &- &- &- &9.5\\
\mcgc   &\third{0.967} &\third{0.959} &\third{0.799} 
        &\third{0.867}  &\third{\revision{0.967}}             
        &0.860 &0.874  &0.596 &0.597 &\revision{0.860}
        &\third{0.567} &0.545 &0.164 &0.186 &\revision{0.553}
        &0.902 &0.895  &0.716 &0.771 &\revision{0.902} &\third{\revision{4.6}}\\ 
\mvagc  &0.774 &0.710 &0.267 &0.329 &\revision{0.790}
        &0.907 &0.915 &0.685 &0.720 &\revision{0.907}
        &0.552 &0.462 &0.191 &0.201 &\revision{0.549}
        &0.874 &0.866 &0.650 &0.708 &\revision{0.874} &5.5\\ 
\magc   &0.714 &0.451 &0.040 &0.030 &\revision{0.714}
        &0.564 &0.520 &0.413 &0.315 &\revision{0.565}
        &0.484 &0.424 &0.057 &0.062 &\revision{0.485}
        &\third{0.928} &\third{0.923} &\third{0.771} &\third{0.827} &\third{\revision{0.928}} &\revision{7.4}\\ \cline{1-17}
\ours   &\second{0.978} &\second{0.974} &\second{0.830} 
        &\second{0.911} &\second{\revision{0.978}}  
        &\second{0.927} &\second{0.930} &\second{0.727} &\second{0.779}	&\second{\revision{0.927}} 
        &{0.559} &0.455	&\first{0.211}	&\first{0.223}	&\third{\revision{0.558}}
        &\first{0.934}  &\first{0.930}  &\first{0.789} &\first{0.841} &\first{\revision{0.933}}	&\first{1.7}\\
\qours  &\first{1.000} &\first{1.000} &\first{1.000} 
        &\first{1.000}	&\first{\revision{1.000}}      
        &\first{0.930} &\first{0.934} &\first{0.740} &\first{0.786}	&\first{\revision{0.930}}
        &0.554	&0.450 &\second{0.210}	&\second{0.220}	 &\revision{0.555}
        &\second{0.930} &\second{0.925} &\second{0.775} &\second{0.831}   &\second{\revision{0.930}} &\second{\revision{2.0}}  \\\hline  

\multirow{2}{*}{\bf{Method} } & \multicolumn{5}{c|}{\bf{Amazon photos}} & \multicolumn{5}{c|}{\bf{Amazon computers}} & \multicolumn{5}{c|}{\bf{MAG-eng}} & \multicolumn{5}{c|}{\bf{MAG-phy}} &\multirow{2}{*}{\shortstack{\bf{Overall}\\ \bf{rank}}} \\ \cline{2-21}
&Acc  &F1 & NMI  & ARI &\revision{Purity} &Acc &F1 & NMI  & ARI &\revision{Purity} &Acc &F1 &NMI  & ARI  &\revision{Purity} &Acc  &F1 & NMI  & ARI &\revision{Purity} & \\ \hline
\wmsc    &0.323	&0.285	&0.152	&0.103 &\revision{0.392}
        &0.248	&0.191	&0.165	&0.090 &\revision{0.375} &- &- &- &- &- &- &- &- &- &- &9.7\\
\twocmv &0.523  &0.434  &0.450 &0.332 &\revision{0.638} 
        &0.309	&0.269  &0.312	&0.135 &\revision{0.524}
        &- &- &- &- &- &- &- &- &- &- &6.8\\ 
\mega   &0.328  &0.292  &0.265  &0.043 &\revision{0.427}
        &0.319  &0.170  &0.230	&0.081 &\revision{0.483}
        &- &- &- &- &- &- &- &- &- &- &\revision{8.3}\\ 
\hdmi   &0.273 &0.126 &0.026 &0.018 &\revision{0.289}
        &0.303 &0.111 &0.034 &0.027	&\revision{0.375}
        &- &- &- &- &- &- &- &- &- &-&8.5\\
\uramn  &0.669 &\third{0.642} &0.521 &0.427 
        &\revision{0.689}
        &0.353 &0.280 &0.364 &0.163 &\revision{0.588}	
        &- &- &- &- &- &- &- &- &- &-&\revision{6.0}\\
\omac   &0.307 &0.153  &0.087 &0.012 &\revision{0.298}
        &0.340 &0.100  &0.020 &0.034 &\revision{0.380}
        &- &- &- &- &- &- &- &- &- &-&\revision{9.5}\\
\dmg    &0.603 &0.548 &0.508 &0.391 &\revision{0.671} 
        &0.401 &0.295 &0.384 &0.200  &\revision{0.598}
        &- &- &- &- &- &- &- &- &- &-&6.3\\
\lmgec  &0.626 &0.606 &0.530 &0.423 &\revision{0.703}
        &0.410 &0.304 &0.374 &0.240  &\revision{0.614}
        &- &- &- &- &- &- &- &- &- &-&\revision{5.1}\\  
\magcn  &0.528 &0.454 &0.456 &0.314 &\revision{0.587}
&- &- &- &- &- &- &- &- &- &- &- &-  &- &- &- &9.5\\
\mcgc   &\third{0.674} &0.582 &\third{0.595} 
        &\third{0.449} &\third{\revision{0.754}}
        &\third{0.569} &\third{0.501} &\third{0.557} &\third{0.419} &\third{\revision{0.726}}  &- &- &- &- &- &- &- &- &- &- &\third{\revision{4.6}}\\ 
\mvagc  &0.615 &0.568 &0.558 &0.384 &\revision{0.726}
        &0.516 &0.426 &0.512 &0.365 &\revision{0.697}
        &\third{0.256} &\third{0.108} &\third{0.355} &\third{0.139} &\third{\revision{0.293}}
        &\third{0.314} &\third{0.107} &\third{0.238} &\third{0.022} &\third{\revision{0.376}} &5.5\\ 
\magc   &0.646 &0.571 &0.591 &0.384 &\revision{0.687}
        &0.447 &0.438 &0.323 &0.158 &\revision{0.481} 
        &- &- &- &- &- &- &- &- &- &- &\revision{7.4}\\ \hline
\ours   &\first{0.786}  &\first{0.710}  &\first{0.670}	 
        &\first{0.622}	&\first{\revision{0.819}}    
        &\second{0.585}	&\second{0.507} &\first{0.589}	&\first{0.441}	&\second{\revision{0.740}}      
        &\first{0.464} &\first{0.369}  &\first{0.575}  &\first{0.332}	&\first{\revision{0.597}}     
        &\first{0.582}	&\first{0.455}  &\first{0.620}	&\first{0.449}	&\first{\revision{0.725}} &\first{1.7}\\ 
\qours  &\second{0.782} &\second{0.705} &\second{0.657}   
        &\second{0.618}	&\second{\revision{0.815}}        
        &\first{0.604}	&\first{0.515}  &\second{0.577}	&\second{0.426}	 &\first{\revision{0.744}}   
        &\second{0.455}  &\second{0.352} &\second{0.570} &\second{0.329}	 &\second{\revision{0.583}}  
        &\second{0.561}	&\second{0.413} &\second{0.608}	&\second{0.439} &\second{\revision{0.704}} &\second{\revision{2.0}}\\  \hline
\end{tabular}
}
\end{small}
\label{tab:revise_clu1}
\vspace{-3mm}
\end{table*}

\stitle{Baselines} 
For embedding, we compare with 8 baselines, including 3 attributed network embedding methods \pane \cite{yang2020scaling}, \aneci \cite{liu2022robust} and \conn \cite{tanCollaborativeGraphNeural2024} that are applied to a multi-view attributed graph by aggregating the graph adjacency matrices and concatenating the attribute views, 3 attributed multiplex graph embedding methods \omac~\cite{fanOne2MultiGraphAutoencoder2020}, \hdmi~\cite{jingHDMIHighorderDeep2021} 
and \uramn~\cite{Zhang2022UnsupervisedRL} that are applied to a multi-view attributed graph by concatenating the attribute views when necessary, and 2 multi-view attributed graph embedding methods \dmg \cite{Mo2023DisentangledMG} and \lmgec \cite{fettalSimultaneousLinearMultiview2023}.
For clustering, we compare with 12 baselines, including 8 multi-view attributed graph clustering approaches, namely \wmsc\cite{zongWeightedMultiViewSpectral2018}, \mega \cite{whang2020mega} adapted for unsupervised clustering, \twocmv \cite{luong2020novel}, \lmgec, \magcn\cite{chengMultiviewAttributeGraph2021}, \mcgc~\cite{panMultiviewContrastiveGraph2021}, \mvagc~\cite{linGraphFilterbasedMultiview2021a}, and \magc~\cite{linMultiViewAttributedGraph2023}, and 4 embedding methods  \hdmi, \uramn, \omac, and \dmg coupled with spectral clustering.

\input{figs/revise/clu_time}

\revision{\stitle{Implementation}}
Across all datasets, \ours and \qours adopt the same parameter settings $\gamma=0.5$, $\tol=0.001$, $\tmax=50$, and $\ralpha=0.05$. We also conduct experiments to vary parameters.
\revision{We set $K=10$ for KNN graphs by default. For Yelp and IMDB, we use $K=200$ and 500, respectively, since their attribute views are more informative. A larger $K$ incorporates more attribute similarity connections in the KNN graphs.}
Source codes of all competitors are obtained from the respective authors, each tuned with parameters suggested in the respective paper.
We fix the embedding dimension to 64. 
Experiments are conducted on a Linux computer with Intel Xeon 6226R CPU, RTX3090 GPU, and 384 GB RAM. A maximum of 16 CPU threads are available.
Note that \conn, \hdmi, \uramn, \dmg, and \lmgec are GPU-powered, while the other methods, including our \ours and \qours, run on CPU.

\vspace{-0.5mm}
\stitle{Evaluation Settings} \label{sec:evaluation-settings}
\revision{The clustering quality is measured by accuracy (Acc), average per-class macro-F1 score (F1), normalized mutual information (NMI), adjusted Rand index (ARI), and Purity score with respect to ground truth.
ARI ranges from $-0.5$ to 1, whereas the other 4 metrics are in range $[0,1]$.}
The node embedding for classification is evaluated by  Macro-F1 (MaF1) and Micro-F1 (MiF1). 
For all these metrics, a larger value indicates better performance.
Efficiency is measured by the total running time in seconds.
Results are averaged over 5 repeated runs. 
\revision{In Tables \ref{tab:revise_clu1} and \ref{tab:revise_emb}, a ` - ' indicates the method cannot produce results within one day or runs out of memory.}

\begin{table*}[!t]
\centering
\renewcommand{\arraystretch}{1.1}
\caption{Embedding performance for node classification (the top 3 are in blue; darker shades indicate better results).}\vspace{-1mm}

\resizebox{1.0\textwidth}{!}{
\setlength{\tabcolsep}{4pt}
\begin{tabular}
{|=c|+c+c|+c+c|+c+c|+c+c|+c+c|+c+c|+c+c|+c+c|+c|}
\hline
\multirow{2}{*}{\bf{Method}} & \multicolumn{2}{c|}{\bf{RM}} & \multicolumn{2}{c|}{\bf{Yelp}} & \multicolumn{2}{c|}{\bf{IMDB}} & \multicolumn{2}{c|}{\bf{DBLP}} & \multicolumn{2}{c|}{\bf{Amazon photos}} & \multicolumn{2}{c|}{\bf{Amazon computers}} & \multicolumn{2}{c|}{\bf{MAG-eng}} & \multicolumn{2}{c|}{\bf{MAG-phy}}& \multirow{2}{*}{\shortstack{\bf{Overall}\\ \bf{rank}}} \\ \cline{2-17}
&MaF1 & MiF1 &MaF1 & MiF1 &MaF1 & MiF1 &MaF1 & MiF1 &MaF1 & MiF1 &MaF1 & MiF1 &MaF1 & MiF1 &MaF1 & MiF1 &\\ \hline
\pane &\third{0.738} &\third{0.778} &0.904 &0.902 &0.479 &0.494 &0.636 &0.763 &0.783 &0.847 &0.556 &0.674 &\third{0.550} &\third{0.672} &\third{0.547} &\third{0.674} &6.0\\
\aneci &0.539 &0.734 &0.778 &0.826 &0.589 &0.596 &0.880 &0.894 &\third{0.899} &\third{0.915} &0.807 &0.846 &- &- &- &- &6.1\\
\conn &0.569 &0.751 &0.932 &0.926 &\third{0.657} &\third{0.657} &0.725 &0.758 &0.892 &0.914 &\third{0.827} &\third{0.850} &- &- &- &- &\third{4.6}\\
\hdmi &0.446 &0.666 &0.926 &0.918 &0.641 &0.642 &0.916 &0.922 &0.724 &0.792 &0.500 &0.721 &- &- &- &- &5.7\\
\uramn &0.496 &0.690 &0.916 &0.907 &0.640 &0.653 &0.897 &0.905 &0.580 &0.727 &0.313 &0.651 &- &- &- &- &6.7\\
\omac &0.689 &0.745 &0.898 &0.894 &\third{0.657} &\third{0.657} &0.909 &0.915 &0.672 &0.721 &0.442 &0.606 &- &- &- &- &6.1\\
\dmg &0.692 &0.737 &0.902 &0.891 &0.618 &0.624 &\third{0.928} &\third{0.933} &0.796 &0.874 &0.629 &0.757 &- &- &- &- &5.2\\
\lmgec &0.417 &0.717 &\third{0.938} &\third{0.932} &0.597 &0.608 &0.916 &0.922 &0.630 &0.723 &0.347 &0.669 &- &- &- &- &6.2\\ \hline
\ours &\second{0.835} &\second{0.904} &\second{0.941} &\second{0.936} &\second{0.688} &\second{0.687} &\first{0.934} &\first{0.938} &\first{0.918} &\first{0.933} &\first{0.893} &\first{0.907} &\second{0.574} &\second{0.736} &\first{0.702} &\first{0.830} &\first{1.5}\\
\qours &\first{0.856} &\first{0.918} &\first{0.942} &\first{0.937} &\first{0.705} &\first{0.704} &\second{0.932} &\second{0.937} &\second{0.912} &\second{0.929} &\second{0.880} &\second{0.901} &\first{0.588} &\first{0.741} &\second{0.696} &\second{0.827} &\first{1.5}\\\hline

\end{tabular}
}
\label{tab:revise_emb}
\vspace{-3mm}
\end{table*}

\subsection{Effectiveness and Efficiency on Clustering}\label{sec:expcluster}

\ours and \qours generate $\LS$ which is then used as the input of spectral clustering as described in Section \ref{sec:problem}. 
In Table \ref{tab:revise_clu1}, we report the Acc, F1, NMI, ARI, Purity and the averaged overall rank of each method over all 8 datasets across the four metrics.
We highlight the top-$3$ best results on each dataset in blue with darker shades indicating better performance.

\stitle{Effectiveness}
As shown in the last column of Table \ref{tab:revise_clu1}, \ours and \qours achieve the best ranks 1.7 and 2.0 respectively over all 8 datasets, significantly outperforming the best competitor with rank 4.6.
Specifically, for the 5 metrics,  \ours and \qours  achieve better performance compared to existing methods on almost all datasets, except Acc, F1 and Purity on IMDB.
For the RM, Yelp, IMDB, and DBLP datasets in Table \ref{tab:revise_clu1}, \ours and \qours achieve improvements in NMI over the best baseline  by up to 20.1\%, 1.5\%, 1.6\% and 1.8\%, respectively.
On Amazon photos, \ours and \qours outperform the best competitor \mcgc by large margins up to 11.2\% in Acc, 12.8\% in F1, 7.5\% in NMI, 17.3\% in ARI and 6.5\% in Purity. 
Remarkably, on the large-scale MAG-eng and MAG-phy datasets in Table \ref{tab:revise_clu1}, \ours and \qours surpass the only scalable baseline \mvagc by up to 23.8\% in Acc, 30.5\% in F1, 30.1\% in NMI, 31.0\% in ARI and 32.7\% in Purity on average.
The results in Table \ref{tab:revise_clu1} validate the effectiveness of the objective formulated in Section \ref{sec:sgfobjective} and the techniques in \ref{sec:algorithms} to solve it, which aligns the spectrum of $\LS$ with the community and connectivity properties.
On the other hand, \mcgc aligns a unified graph with each view but overlooks its intrinsic structure, leading to inferior performance.
The results in Table \ref{tab:revise_clu1}  underscore that 
\ours and \qours effectively generate a \mvag Laplacian $\LS$ that reliably reveals the underlying clusters in real-world multi-view attributed graphs. %

\stitle{Efficiency}
For our methods, we record the \textit{total time cost} of computing view Laplacians from $\GS$, running \ours or \qours, and performing clustering.
Figure \ref{fig:revise_clu_time} displays the running time of all methods, with the competitor delivering the best clustering quality marked by a star for each dataset. 
The $y$-axis is  time in seconds on a logarithmic scale. 
Regardless of clustering quality, \ours and \qours consistently demonstrate leading efficiency across all datasets except RM. Compared to the marked competitors, \ours and \qours are often faster by orders of magnitude. For instance, on DBLP, the best baseline \magc requires 35.98 seconds to finish, while \ours and \qours only take 2.008 and 0.788 seconds, respectively, achieving a 17.9$\times$ and 45.7$\times$ speedup. On Amazon photos, the marked baseline \mcgc requires 5102 seconds, whereas our methods \ours and \qours take only 1.465 and 1.129 seconds, attaining a significant speedup of over 3000$\times$.
On MAG-eng and MAG-phy, where most baselines run out of memory or cannot finish within one day, our methods achieve the highest efficiency with the best quality. 
Our methods demonstrate a huge speedup over baselines with quadratic complexity, \eg \magc, and the GNN models that are expensive to train, \eg \hdmi.
Furthermore, \qours is consistently faster than \ours on all datasets.
For example, \ours requires 1206 and 1970 seconds for clustering MAG-eng and MAG-phy, respectively, while \qours requires only 583 and 783 seconds.

Moreover,   
\ours and \qours also exhibit high memory efficiency. On large-scale datasets, our methods only use 18.7 GB for MAG-eng and 32.3 GB for MAG-phy, while \mvagc requires 137 GB and 184 GB, respectively, and all other baselines  are out of memory.
These  results highlight the efficacy of the algorithm designs presented in Section \ref{sec:algorithms}.

\input{figs/revise/emb_time}

\subsection{Effectiveness and Efficiency on Embedding }\label{sec:expembedding}

\ours and $\qours$ generate $\LS$ which is then used as the input for classic network embedding methods as described in Section \ref{sec:problem}. Specifically,  on    large MAG-eng and MAG-phy, the scalable \sketchne~\cite{xieSketchNEEmbeddingBillionScale2023} is utilized, while \netmf~\cite{qiuNetworkEmbeddingMatrix2018a} is used for the remaining datasets. 
We evaluate the embedding quality by node classification.
For each method that outputs embeddings, a logistic regression classifier is trained on 20\% of the ground truth class labels (1\% for   MAG-eng and MAG-phy), with the remaining labels used for testing.
In Table \ref{tab:revise_emb}, we report the results of classification performance (MaF1, MiF1). Figure \ref{fig:revise_emb_time} compares the efficiency of all methods, measured by total embedding time in seconds. 

\stitle{Effectiveness}
In Table \ref{tab:revise_emb}, our methods \ours and \qours consistently claim the top two places, with the best overall rank 1.5 over all metrics on all datasets, significantly higher than 
the best competitor with overall rank 4.6. 
For example, on IMDB in Table \ref{tab:revise_emb}, \qours and \ours take the first and second places respectively, surpassing \omac and \conn, which rank third, by up to 4.8\% in Macro-F1 and 4.7\% in Micro-F1. On the Amazon computers dataset, our methods outperform the runner-up \conn by up to 6.6\% in Macro-F1 and 5.7\% in Micro-F1.
Moreover, on the large datasets, MAG-eng and MAG-phy, our methods \qours and \ours surpass \pane, the sole competitor with sufficient scalability, proving that our methods are more effective in producing high-quality embeddings for these large-scale datasets.
{Compared with sophisticated GNN methods focused on the common and specific aspects of each view, \eg \dmg, our methods better preserve structural properties, thus allowing classic methods to produce high-quality node embeddings for \mvags.}
The results in Table \ref{tab:revise_emb} confirm the effectiveness of the objective and techniques developed in Section \ref{sec:sgfobjective} and \ref{sec:algorithms}.

\stitle{Efficiency}
Figure \ref{fig:revise_emb_time} displays the \textit{total time cost} for \mvag embedding on 8 datasets, with the competitors delivering the best embedding quality marked for each dataset. The $y$-axis represents running time in seconds on a logarithmic scale. \qours achieves the best efficiency on all datasets, often outperforming the compared methods by orders of magnitude; \ours is also faster than all baselines except on the smallest dataset, RM. {For instance, in Figure \ref{fig:time-emb-imdb}, while \conn and \omac are the runner-ups after \ours and \qours in quality, our methods are up to 222$\times$ and $489\times$ faster, respectively.} On the million-scale datasets MAG-eng and MAG-phy, \qours requires 555 and 939 seconds each, achieving 32.1$\times$ and 71.6$\times$ speedup over the only scalable baseline \pane. Moreover, \qours is faster than \ours on all datasets. 
For memory usage,  our methods also achieve high space efficiency. For example, to produce embeddings for large-scale datasets MAG-eng and MAG-phy, both \ours and \qours leave a memory footprint of 61.6 GB  and 95.4 GB, respectively. However,  the only scalable baseline \pane requires 221 GB and 299 GB, while all other methods run out of memory. 
These  results highlight the efficacy of the algorithm designs presented in Section \ref{sec:algorithms}.

\subsection{Experimental Analysis} \label{sec:ablation-study}

\begin{figure}[!t]
\vspace{-4mm}
\definecolor{darkgray176}{RGB}{176,176,176}
\definecolor{darkorange25512714}{RGB}{255,127,14}
\definecolor{forestgreen4416044}{RGB}{44,160,44}
\definecolor{lightgray204}{RGB}{204,204,204}
\definecolor{steelblue31119180}{RGB}{31,119,180}
\definecolor{tabpurple}{RGB}{148, 103, 189}
\definecolor{tabbrown}{RGB}{140, 86, 75}
\definecolor{tabpink}{RGB}{227, 119, 194}

\subfloat[Yelp]{
\begin{tikzpicture}[scale=0.9,every mark/.append style={mark size=2pt}]
    \begin{axis}[
    height=\columnwidth/2.5,
    width=\columnwidth/1.9,
xlabel={\em t},
every axis x label/.style={font=\footnotesize,at={{(0.68,0.0)}},right=12mm,below=1mm},
legend cell align={left},
legend style={
  at={(1.05,0.30)},
  anchor=south east,
  draw=none,
  fill=none,
  font=\footnotesize
},
axis lines=left,
tick align=center,
tick pos=left,
x grid style={darkgray176},
xmin=-1.15, xmax=51.15,
xtick style={color=black},
y grid style={darkgray176},
ymin=0.50, ymax=1.05,
ytick style={color=black},
yticklabel style = {font=\footnotesize,xshift=0.5ex},
xticklabel style = {font=\footnotesize},
]

\addplot [semithick, line width=1.2pt, tabpurple]
plot coordinates {
(0,	0.760536472)
(1,	0.780923707)
(2,	0.783980054)
(3,	0.708717996)
(4,	0.667449465)
(5,	0.688549779)
(6,	0.642046288)
(7,	0.664041259)
(8,	0.647235563)
(9,	0.640292227)
(10,	0.653410539)
(11,	0.635583035)
(12,	0.632815947)
(13,	0.637640451)
(14,	0.632162342)
(15,	0.629532884)
(16,	0.627621929)
(17,	0.630580282)
(18,	0.617471376)
(19,	0.63136519)
(20,	0.628895458)
(21,	0.629695979)
(22,	0.629855655)
(23,	0.629313491)
(24,	0.629536396)
(25,	0.62956722)
(26,	0.618039492)
(27,	0.629488837)
(28,	0.629360505)
(29,	0.629488317)
(30,	0.62948497)
(31,	0.629448907)
(32,	0.629424199)
(33,	0.629481706)
(34,	0.629488516)
(35,	0.629300054)
(36,	0.629485235)
(37,	0.629447335)
(38,	0.629422879)
(39,	0.629485526)
(40,	0.629467834)
(41,	0.62931763)
(42,	0.629380199)
(43,	0.617868855)
(44,	0.629485935)
(45,	0.629392757)
(46,	0.629485641)
(47,	0.629485871)
(48,	0.629139698)
(49,	0.629284889)
(50,	0.62948567)
};

\addlegendentry{$h(\wv)$}
\addplot [semithick, forestgreen4416044,line width=1.2pt]
plot coordinates {
(0,	0.695485845)
(1,	0.674062739)
(2,	0.643458301)
(3,	0.914307575)
(4	,0.931522571)
(5	,0.932287682)
(6	,0.931522571)
(7	,0.919280796)
(8	,0.932670237)
(9	,0.930374904)
(10	,0.924636572)
(11	,0.930374904)
(12	,0.93075746)
(13	,0.933052793)
(14	,0.928844682)
(15	,0.928462127)
(16	,0.928844682)
(17	,0.927314461)
(18	,0.930374904)
(19	,0.927314461)
(20	,0.929227238)
(21	,0.928844682)
(22	,0.928079572)
(23	,0.928844682)
(24	,0.928462127)
(25	,0.928462127)
(26	,0.928462127)
(27	,0.928462127)
(28	,0.928462127)
(29	,0.928462127)
(30	,0.928462127)
(31	,0.928462127)
(32	,0.928462127)
(33	,0.928462127)
(34	,0.928462127)
(35	,0.928462127)
(36	,0.928462127)
(37	,0.928462127)
(38	,0.928462127)
(39	,0.928462127)
(40	,0.928462127)
(41	,0.928462127)
(42	,0.928462127)
(43	,0.928462127)
(44	,0.928462127)
(45	,0.928462127)
(46	,0.928462127)
(47	,0.928462127)
(48	,0.928462127)
(49	,0.928462127)
(50	,0.928462127)
};
\addlegendentry{Acc}
\addplot[mark=|,mark size=1pt,color=gray] coordinates {(28,0.880585915)};
\addplot[mark=|,mark size=1pt,color=gray] coordinates {(28,0.820585915)};
\addplot[mark=|,mark size=1pt,color=gray] coordinates {(28,0.760585915)};
\addplot[mark=|,mark size=1pt,color=gray] coordinates {(28,0.70585915)};
\addplot[mark=|,mark size=1pt,color=gray] coordinates {(28,0.640585915)};
\addplot[mark=|,mark size=1pt,color=gray] coordinates {(28,0.580585915)};
\addplot[mark=|,mark size=1pt,color=gray] coordinates {(28,0.520585915)};
\addplot[mark=*,mark size=1.5pt,color=black] coordinates {(28,0.629360505)};
\addplot[mark=*,mark size=1.5pt,color=black] coordinates {(28,0.928462127)};
\end{axis}
\end{tikzpicture}\hspace{-1mm}
}
\subfloat[IMDB]{
\begin{tikzpicture}[scale=0.9,every mark/.append style={mark size=2pt}]
    \begin{axis}[
    height=\columnwidth/2.5,
    width=\columnwidth/1.9,
legend cell align={left},
legend style={
  at={(1.05,0.55)},
  anchor=south east,
  draw=none,
  fill=none,
  font=\footnotesize
},
xlabel={\em t},
every axis x label/.style={font=\footnotesize,at={{(0.68,0.0)}},right=12mm,below=1mm},
axis lines=left,
tick align=center,
tick pos=left,
x grid style={darkgray176},
xmin=-1.15, xmax=51.15,
xtick style={color=black},
xtick={0,20,40},
xticklabels ={0,20,40},
y grid style={darkgray176},
ymin=0.25, ymax=1.05,
ytick style={color=black},
yticklabel style = {font=\footnotesize,xshift=0.5ex},
xticklabel style = {font=\footnotesize},
]
\addplot [semithick, line width=1.2pt, tabpurple]
plot coordinates {
(0	,   0.826433102)
(1	,   0.877969367)
(2	,   0.877945406)
(3	,   0.760437323)
(4	,   0.706408793)
(5	,   0.669078318)
(6	,   0.68537366)
(7	,   0.663375455)
(8	,   0.668189677)
(9	,   0.685350205)
(10,   	0.665357317)
(11,   	0.686828995)
(12,   	0.664943361)
(13,   	0.664031068)
(14,   	0.672186084)
(15,   	0.665458892)
(16,   	0.668935484)
(17,   	0.661635883)
(18,   	0.661619688)
(19,   	0.66204012)
(20,   	0.668380273)
(21,   	0.666267596)
(22,   	0.668240376)
(23,   	0.668104077)
(24,   	0.668236268)
(25,   	0.668153147)
(26,   	0.662498863)
(27,   	0.666984689)
(28,   	0.668034207)
(29,   	0.667740514)
(30,   	0.668049958)
(31,   	0.6679691)
(32,   	0.668167897)
(33,   	0.66522587)
(34,   	0.668201456)
(35,   	0.665068427)
(36,   	0.668220803)
(37,   	0.662884086)
(38,   	0.667292897)
(39,   	0.66684042)
(40,   	0.665383034)
(41,   	0.668203508)
(42,   	0.663547776)
(43,   	0.665213329)
(44,   	0.662001165)
(45,   	0.661770493)
(46,   	0.668221329)
(47,   	0.665449412)
(48,   	0.667931706)
(49,   	0.668222937)
(50,	0.668088169)
};
\addlegendentry{$h(\wv)$}
\addplot [semithick, forestgreen4416044,line width=1.2pt]
plot coordinates {
(0	,0.377183099)
(1	,0.377183099)
(2	,0.377183099)
(3	,0.376619718)
(4	,0.378028169)
(5	,0.556619718)
(6	,0.554647887)
(7	,0.556056338)
(8	,0.550985915)
(9	,0.383380282)
(10,	0.556338028)
(11,	0.381971831)
(12,	0.556619718)
(13,	0.556901408)
(14,	0.513239437)
(15,	0.552676056)
(16,	0.545070423)
(17,	0.554647887)
(18,	0.551830986)
(19,	0.550140845)
(20,	0.548450704)
(21,	0.550704225)
(22,	0.547887324)
(23,	0.550422535)
(24,	0.550422535)
(25,	0.550704225)
(26,	0.550704225)
(27,	0.550985915)
(28,	0.548450704)
(29,	0.550985915)
(30,	0.552112676)
(31,	0.550422535)
(32,	0.549295775)
(33,	0.551267606)
(34,	0.550704225)
(35,	0.550704225)
(36,	0.550704225)
(37,	0.550704225)
(38,	0.551549296)
(39,	0.550985915)
(40,	0.552394366)
(41,	0.550985915)
(42,	0.549295775)
(43,	0.549577465)
(44,	0.550422535)
(45,	0.550704225)
(46,	0.551267606)
(47,	0.549014085)
(48,	0.551549296)
(49,	0.550704225)
(50,	0.551830986)
};
\addlegendentry{Acc}
\addplot[mark=|,mark size=1pt,color=gray] coordinates {(24,0.590985915)};
\addplot[mark=|,mark size=1pt,color=gray] coordinates {(24,0.510985915)};
\addplot[mark=|,mark size=1pt,color=gray] coordinates {(24,0.430985915)};
\addplot[mark=|,mark size=1pt,color=gray] coordinates {(24,0.350985915)};
\addplot[mark=|,mark size=1pt,color=gray] coordinates {(24,0.270985915)};
\addplot[mark=*,mark size=1.5pt,color=black] coordinates {(24,0.66684042)};
\addplot[mark=*,mark size=1.5pt,color=black] coordinates {(24,0.550985915)};
\end{axis}
\end{tikzpicture}\hspace{-3mm}
}
\vspace{-1mm}
\caption{{{Varying  number of iterations $t$ in  \ours for clustering accuracy; $\bullet$ marks when  termination condition is met). }}} \label{fig:converge}
\vspace{-4mm}
\end{figure}
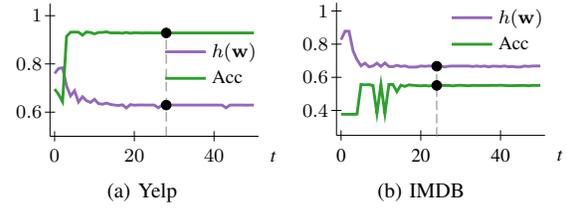

\vspace{-1mm}
\stitle{Convergence evaluation and $\tmax$} When varying the number of iterations $t$ in \ours, Figure \ref{fig:converge} shows the convergence of the objective $h(\wv)$ that decreases and then becomes stable, while the corresponding clustering accuracy (Acc) improves. The black dots mark the iteration when the termination condition by $\epsilon$ at Line 7 of Algorithm \ref{alg:direct} is met.
Observe that $h(\wv)$ usually converges before termination. Thus, we set $\tmax=50$ by default at Line 2 of Algorithm \ref{alg:direct}.

\definecolor{darkblue}{RGB}{0,0,139}
\definecolor{darkgray176}{RGB}{176,176,176}
\definecolor{darkgreen}{RGB}{0,100,0}
\definecolor{darkorange25512714}{RGB}{255,127,14}
\definecolor{darkred}{RGB}{139,0,0}
\definecolor{forestgreen4416044}{RGB}{44,160,44}
\definecolor{goldenrod}{RGB}{218,165,32}
\definecolor{lightgray204}{RGB}{204,204,204}
\definecolor{lightseagreen0171189}{RGB}{0,171,189}
\definecolor{lightsteelblue161199224}{RGB}{161,199,224}
\definecolor{magenta}{RGB}{255,0,255}
\definecolor{steelblue31119180}{RGB}{31,119,180}
\definecolor{teal2110129}{RGB}{2,110,129}
\definecolor{myred}{HTML}{fd7f6f}
\definecolor{mywhite}{HTML}{D8D8D8}
\definecolor{myorange}{HTML}{D7191C}
\definecolor{myblue}{HTML}{7eb0d5}
\definecolor{mygreen}{HTML}{b2e061}
\definecolor{mypurple}{HTML}{bd7ebe}
\definecolor{myorange}{HTML}{ffb55a}
\definecolor{myyellow}{HTML}{ffee65}
\definecolor{mypurple2}{HTML}{beb9db}
\definecolor{mypink}{HTML}{fdcce5}
\definecolor{mycyan}{HTML}{8bd3c7}
\definecolor{mycyan2}{HTML}{00ffff}
\definecolor{myblue2}{HTML}{115f9a}
\definecolor{myred2}{HTML}{c23728}
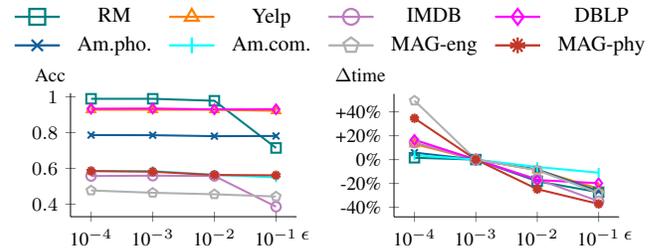
\begin{figure}[!t]
\centering
\begin{small}
\begin{tikzpicture}
    \begin{customlegend}[legend columns=3,
        legend entries={RM, Yelp, IMDB, DBLP, Am.pho., Am.com., MAG-eng, MAG-phy},
        legend columns=4,
        legend style={at={(0.45,1.35)},anchor=north,draw=none,font=\footnotesize,column sep=0.15cm}]
    \addlegendimage{line width=0.3mm,mark size=3pt,mark=square,color=teal}
    \addlegendimage{line width=0.3mm,mark size=3pt,mark=triangle,color=orange}
    \addlegendimage{line width=0.3mm,mark size=3pt,mark=o,color=mypurple}
    \addlegendimage{line width=0.3mm,mark size=3pt,mark=diamond,color=magenta}
    \addlegendimage{line width=0.3mm,mark size=3pt,mark=x,color=myblue2}
    \addlegendimage{line width=0.3mm,mark size=3pt,mark=|,color=mycyan2}
    \addlegendimage{line width=0.3mm,mark size=3pt,mark=pentagon,color=darkgray176}
    \addlegendimage{line width=0.3mm,mark size=3pt,mark=10-pointed star,color=myred2}
    \end{customlegend}
\end{tikzpicture}
\\[-\lineskip]
\vspace{-4mm}
\subfloat{
\resizebox{0.43\columnwidth}{!}{
\begin{tikzpicture}[scale=0.98,every mark/.append style={mark size=2pt}]
    \begin{axis}[
    height=\columnwidth/2.6,
    width=\columnwidth/1.8,
    axis lines=left,
    anchor = north,
    xmin=0.5, xmax=6,
    ymin=0.33, ymax=1.02,
    xtick={1,2.5,4,5.5},
    xticklabels={10$^{-4}$, 10$^{-3}$, 10$^{-2}$, 10$^{-1}$},
    ytick={0.2,0.4,0.6,0.8,1.0},
    scaled y ticks = false,
    ylabel={Acc},
    xticklabel style = {font=\footnotesize},
    yticklabel style = {font=\footnotesize},
    every axis y label/.style={font=\footnotesize, at={(current axis.north west)},right=-3mm,above=1mm},
    xlabel={$\epsilon$},
    every axis x label/.style={font=\footnotesize,at={{(0.68,0.0)}},right=12mm,below=1mm},
    ]
    \addplot[line width=0.3mm,mark=square,color=teal]  %
        plot coordinates {
    (1, 0.989010989010989)
    (2.5, 0.989010989010989)
    (4, 0.978021978021978)
    (5.5, 0.714285714285714)
        };    
    \addplot[line width=0.3mm,mark=triangle,color=orange] %
    plot coordinates {
    (1,0.9276970160673298)
    (2.5,0.9276970160673298)
    (4,0.9276970160673298)
    (5.5,0.921733063451213)
    };
    \addplot[line width=0.3mm,mark=o,color=mypurple] %
    plot coordinates {
    (1,0.5574647887323944)
    (2.5,0.5591549295774648)
    (4,0.5577464788732395)
    (5.5,0.38591549295774646)
    };
    \addplot[line width=0.3mm,mark=diamond,color=magenta] %
    plot coordinates {
    (1,0.9334483608577767)
    (2.5,0.9346807986196697)
    (4,0.9297510475720976)
    (5.5,0.9304905102292335)

    };
    \addplot[line width=0.3mm,mark=x,color=myblue2] %
    plot coordinates {
    (1,0.7862745098039216)
    (2.5,0.7856209150326797)
    (4,0.7801307189542486)
    (5.5,0.7812418300653677)

    };
    \addplot[line width=0.3mm,mark=|,color=mycyan2] %
    plot coordinates {
    (1,0.585151250727167)
    (2.5,0.585151250727167)
    (4,0.5642233856893584)
    (5.5,0.5496800465386984)

    };
    \addplot[line width=0.3mm,mark=pentagon,color=darkgray176] %
    plot coordinates {
    (1,0.4767264666982077 )
    (2.5,0.4637772784712659)
    (4,0.45534177972410333)
    (5.5,0.442793947018902)

    };
    \addplot[line width=0.3mm,mark=10-pointed star,color=myred2] %
    plot coordinates {
    (1,0.585151250727167)
    (2.5,0.581512507271675)
    (4,0.5631351964914129)
    (5.5,0.56142831168787)

    };
    
    \end{axis}
\end{tikzpicture}\hspace{-3mm}
}
}%
\subfloat{
\resizebox{0.49\columnwidth}{!}{
\begin{tikzpicture}[scale=0.98,every mark/.append style={mark size=2pt}]
    \begin{axis}[
    height=\columnwidth/2.6,
    width=\columnwidth/1.8,
        axis lines=left,
        xmin=0.5, xmax=6,
        ymin=-48, ymax=55,
        xtick={1,2.5,4,5.5},
        ytick={-40,-20,0,20,40},
        xticklabels={10$^{-4}$, 10$^{-3}$, 10$^{-2}$, 10$^{-1}$},
        yticklabels={-40\%,-20\%,0\%,+20\%,+40\%},
        scaled y ticks = false,
        ylabel={$\Delta$time},
        xticklabel style = {font=\footnotesize},
        yticklabel style = {font=\footnotesize},
        every axis y label/.style={font=\footnotesize, at={(current axis.north west)},right=-5mm,above=1mm},
        xlabel={$\epsilon$},
        every axis x label/.style={font=\footnotesize,at={{(0.68,0.0)}},right=12mm,below=1mm},
    ]
    \addplot[line width=0.3mm,mark=square,color=teal]  %
    plot coordinates {
    (1, 1.498127)
    (2.5, 0.000)
    (4, -18.164794)
    (5.5, -27.340824)

    };  
    \addplot[line width=0.3mm,mark=triangle,color=orange] %
    plot coordinates {
    (1,13.012730)
    (2.5,0.000000)
    (4,-8.769448)
    (5.5,-24.328147)
    };
    \addplot[line width=0.3mm,mark=o,color=mypurple] %
    plot coordinates {
    (1,14.270941)
    (2.5,0)
    (4,-16.445183)
    (5.5,-34.883721)
    };
    \addplot[line width=0.3mm,mark=diamond,color=magenta] %
    plot coordinates {
    (1,16.555925)
    (2.5,0)
    (4,-17.131979695)
    (5.5,-19.796954315)
    };
    \addplot[line width=0.3mm,mark=x,color=myblue2] %
    plot coordinates {
    (1,5.830258)
    (2.5,0)
    (4,-8.339483)
    (5.5,-26.494465)
    };
    \addplot[line width=0.3mm,mark=|,color=mycyan2] %
    plot coordinates {
    (1,3.700641)
    (2.5,0)
    (4,-6.181517)
    (5.5,-11.019227)
    };
    \addplot[line width=0.3mm,mark=pentagon,color=darkgray176] %
    plot coordinates {
    (1,49.366285)
    (2.5,0.000000)
    (4,-8.921524)
    (5.5,-27.968215)
    };
    \addplot[line width=0.3mm,mark=10-pointed star,color=myred2] %
    plot coordinates {
    (1,34.595001)
    (2.5,0)
    (4,-24.755607)
    (5.5,-37.141746)
    };
    
    \end{axis}
\end{tikzpicture}\hspace{0mm}
}
}%

\vspace{-1mm}
\end{small}
\caption{Varying $\epsilon$ for \ours.} \label{fig:param-eps}
\vspace{-4mm}
\end{figure}
\definecolor{darkblue}{RGB}{0,0,139}
\definecolor{darkgray176}{RGB}{176,176,176}
\definecolor{darkgreen}{RGB}{0,100,0}
\definecolor{darkorange25512714}{RGB}{255,127,14}
\definecolor{darkred}{RGB}{139,0,0}
\definecolor{forestgreen4416044}{RGB}{44,160,44}
\definecolor{goldenrod}{RGB}{218,165,32}
\definecolor{lightgray204}{RGB}{204,204,204}
\definecolor{lightseagreen0171189}{RGB}{0,171,189}
\definecolor{lightsteelblue161199224}{RGB}{161,199,224}
\definecolor{magenta}{RGB}{255,0,255}
\definecolor{steelblue31119180}{RGB}{31,119,180}
\definecolor{teal2110129}{RGB}{2,110,129}
\definecolor{myred}{HTML}{fd7f6f}
\definecolor{mywhite}{HTML}{D8D8D8}
\definecolor{myorange}{HTML}{D7191C}
\definecolor{myblue}{HTML}{7eb0d5}
\definecolor{mygreen}{HTML}{b2e061}
\definecolor{mypurple}{HTML}{bd7ebe}
\definecolor{myorange}{HTML}{ffb55a}
\definecolor{myyellow}{HTML}{ffee65}
\definecolor{mypurple2}{HTML}{beb9db}
\definecolor{mypink}{HTML}{fdcce5}
\definecolor{mycyan}{HTML}{8bd3c7}
\definecolor{mycyan2}{HTML}{00ffff}
\definecolor{myblue2}{HTML}{115f9a}
\definecolor{myred2}{HTML}{c23728}
\begin{figure}[!t]
\vspace{-2mm}
\centering
\begin{small}
\subfloat{
\resizebox{0.45\columnwidth}{!}{
\begin{tikzpicture}[scale=0.98,every mark/.append style={mark size=2pt}]
    \begin{axis}[
    height=\columnwidth/2.6,
    width=\columnwidth/1.8,
        axis lines=left,
        xmin=0.5, xmax=7.5,
        ymin=0, ymax=1.05,
        xtick={1,2,3,4,5,6,7},
        xticklabel style = {font=\small},
        xticklabels={-2,-1,-0.5,0,0.5,1,2},
        ytick={0,0.2,0.4,0.6,0.8,1.0},
        scaled y ticks = false,
        ylabel={Acc},
        xticklabel style = {font=\footnotesize},
        yticklabel style = {font=\footnotesize},
        every axis y label/.style={font=\footnotesize, at={ (current axis.north west)},right=-3mm,above=1mm},
        xlabel={$\gamma$},
        every axis x label/.style={font=\footnotesize,at={{(0.68,0.0)}},right=12mm,below=1mm},
    ]
    \addplot[line width=0.3mm,mark=square,color=teal]  %
    plot coordinates {
    (1,0.9780219780219)
    (2,0.9780219780219)
    (3,0.9780219780219)
    (4,0.9890109890)
    (5,1)
    (6,0.98901099)
    (7,0.97802197)
    };    
    \addplot[line width=0.3mm,mark=triangle,color=orange] %
    plot coordinates {
    (1,0.906656465)
    (2,0.9070390)
    (3,0.9177505)
    (4,0.92272379)
    (5,0.9296097)
    (6,0.92578423)
    (7,0.90971690)
    };
    \addplot[line width=0.3mm,mark=o,color=mypurple] %
    plot coordinates {
    (1,0.50901408)
    (2,0.50873239)
    (3,0.51070423)
    (4,0.52225352)
    (5,0.55380281)
    (6,0.42478873)
    (7,0.3771831)
    };
    \addplot[line width=0.3mm,mark=diamond,color=magenta] %
    plot coordinates {
    (1,0.5679073206)
    (2,0.5661819078)
    (3,0.569139758)
    (4,0.9227237949)
    (5,0.9295045600)
    (6,0.9324624106)
    (7,0.9297510475)
    };
    \addplot[line width=0.3mm,mark=x,color=myblue2] %
    plot coordinates {
    (1,0.55620915)
    (2,0.55620915)
    (3,0.55620915)
    (4,0.7827451)
    (5,0.78222222)
    (6,0.78640522)
    (7,0.78836601)
    };
    \addplot[line width=0.3mm,mark=|,color=mycyan2] %
    plot coordinates {
    (1,0.46749564)
    (2,0.46749564)
    (3,0.46815008)
    (4,0.58595113)
    (5,0.59962187)
    (6,0.60260325)
    (7,0.58151542)
    };
    \addplot[line width=0.3mm,mark=pentagon,color=darkgray176] %
    plot coordinates {
    (1,0.07381983936327949)
    (2,0.07467767303027659)
    (3,0.09270830264015963)
    (4,0.447634)
    (5,0.454809177875119)
    (6,0.454396661620477)
    (7,0.448723120787321)
    };
    \addplot[line width=0.3mm,mark=10-pointed star,color=myred2] %
    plot coordinates {
    (1,0.152625153143845)
    (2,0.152625153143845)
    (3,0.152625153143845)
    (4,0.5084651800597)
    (5,0.561275380247035)
    (6,0.563365018462223)
    (7,0.545009456260758)
    };
    
    \end{axis}
\end{tikzpicture}\hspace{-2mm}
}
}%
\subfloat{
\resizebox{0.48\columnwidth}{!}{
\begin{tikzpicture}[scale=0.98,every mark/.append style={mark size=2pt}]
    \begin{axis}[
    height=\columnwidth/2.6,
    width=\columnwidth/1.8,
        axis lines=left,
        xmin=0.5, xmax=7.5,
        ymin=0, ymax=1.05,
        xtick={1,2,3,4,5,6,7},
        xticklabel style = {font=\small},
        xticklabels={-2,-1,-0.5,0,0.5,1,2},
        ytick={0,0.2,0.4,0.6,0.8,1.0},
        scaled y ticks = false,
        ylabel={NMI},
        xticklabel style = {font=\footnotesize},
        yticklabel style = {font=\footnotesize},
        every axis y label/.style={font=\footnotesize, at={ (current axis.north west)},right=-3mm,above=1mm},
        xlabel={$\gamma$},
        every axis x label/.style={font=\footnotesize,at={{(0.68,0.0)}},right=12mm,below=1mm},
    ]
    \addplot[line width=0.3mm,mark=square,color=teal]  %
    plot coordinates {
    (1,0.82960858)
    (2,0.82960858)
    (3,0.82960858)
    (4,0.91397155)
    (5,1)
    (6,0.91397155)
    (7,0.82960858)
        };    
    \addplot[line width=0.3mm,mark=triangle,color=orange] %
    plot coordinates {
    (1,0.679058936)
    (2,0.680159313)
    (3,0.6978488828)
    (4,0.71341182)
    (5,0.738781285)
    (6,0.725065277)
    (7,0.675887375)
    };
    \addplot[line width=0.3mm,mark=o,color=mypurple] %
    plot coordinates {
    (1, 0.14241581)
    (2, 0.14218898)
    (3, 0.14499498)
    (4, 0.15417114)
    (5, 0.20929396)
    (6, 0.06716036)
    (7, 0.00242745)
    };
    \addplot[line width=0.3mm,mark=diamond,color=magenta] %
    plot coordinates {
    (1, 0.23762527)
    (2, 0.23576415)
    (3, 0.23917219)
    (4, 0.7134118275)
    (5, 0.7747282598)
    (6, 0.7835851876)
    (7, 0.77541757739)
    };
    \addplot[line width=0.3mm,mark=x,color=myblue2] %
    plot coordinates {
    (1, 0.54119444)
    (2, 0.54119444)
    (3, 0.54119444)
    (4, 0.67854555)
    (5, 0.65702831)
    (6, 0.67225057)
    (7, 0.67765034)
    };
    \addplot[line width=0.3mm,mark=|,color=mycyan2] %
    plot coordinates {
    (1, 0.25620485)
    (2, 0.25620485)
    (3, 0.27142338)
    (4, 0.58906702)
    (5, 0.58134639)
    (6, 0.58099025)
    (7, 0.56310328)
    };
    \addplot[line width=0.3mm,mark=pentagon,color=darkgray176] %
    plot coordinates {
    (1,0.0220936509)
    (2,0.0189911927)
    (3,0.066719922)
    (4,0.563389)
    (5,0.5669034421)
    (6,0.568200438)
    (7,0.562308329)
    };
    \addplot[line width=0.3mm,mark=10-pointed star,color=myred2] %
    plot coordinates {
    (1,0.0117482008)
    (2,0.0117482008)
    (3,0.0117482008)
    (4,0.5581155278677)
    (5,0.6077686194945)
    (6,0.6126048372462)
    (7,0.6041243624401)
    
    };
    \end{axis}
\end{tikzpicture}\hspace{2mm}
}
}%
\end{small}

\vspace{-1mm}
\caption{Varying $\gamma$   for \qours. } \label{fig:param-gamma}
\vspace{-3mm}
\end{figure}
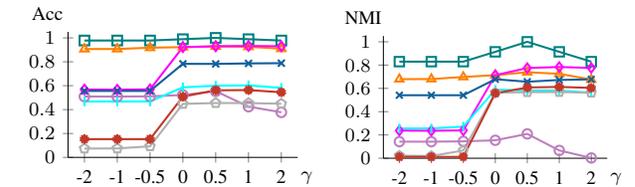

\stitle{Varying $\epsilon$}  We vary parameter $\epsilon$ in the termination condition from $10^{-4}$ to $10^{-1}$ (from tight to loose) and report the results of \ours in Figure \ref{fig:param-eps}. Compared to the default setting of $\epsilon=10^{-3}$, $\Delta$time denotes the ratio of change in running time.
As $\epsilon$ becomes loose from $10^{-4}$ to $10^{-1}$, the clustering quality (Acc) is stable first and then decreases. 
On the other hand, as $\epsilon$ becomes tight, e.g., from $10^{-3}$ to $10^{-4}$, the running time increases significantly ($\Delta\text{time}$) but the quality (Acc) maintains. Thus, we set $\epsilon=0.001$ by default.

\stitle{Varying $\gamma$} 
Parameter $\gamma$ is the coefficient of the regularization term in \eqref{eq:fullobj}. A negative $\gamma$ promotes weight allocation to focus on a single view, while a positive $\gamma$ mitigates such situations and tends to assign similar weights across views. We vary $\gamma$ from $-2$ to 2 and report the accuracy and NMI scores in Figure \ref{fig:param-gamma}.
As $\gamma$ increases from $-2$ to 0.5, the accuracy and NMI of \qours remain relatively stable on Yelp and show noticeable improvement on other datasets.
However, when $\gamma$ varies from 0.5 to 2, the accuracy and NMI degrade on IMDB and RM, while remaining stable on other datasets.
Based on these observations, we set $\gamma=0.5$ by default.

\input{figs/revise/revise_sample}
\stitle{Varying the number of weight vector samples in \qours} 
In Section \ref{sec:pi}, \qours uses $(\nv+1)$ sampled weight vectors by default. 
We vary the number of samples and change $(r+1)$ by $\Delta s \in \{-2,-1,0,+2,+5,+10,+20\}$, and report the results of \qours in Figure \ref{fig:param-sample}, where the left $y$-axis is for clustering Acc and NMI, and the right $y$-axis is for running time.
The removed (resp. added) samples are randomly selected (resp. generated).
Observe that when the number of samples changes by $\Delta s$ from -2 to 0, Acc and NMI scores increase and then become stable afterward with larger delta values. Meanwhile, the time increases significantly due to more expensive objective evaluations to be performed. 
The results in Figure \ref{fig:param-sample} indicate that $(r+1)$ samples are sufficient in practice.

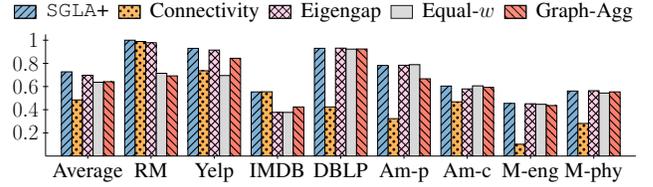
\begin{figure}[!t]
    \centering
    \resizebox{0.96\columnwidth}{!}{%
    \begin{tikzpicture}

\definecolor{darkgray176}{RGB}{176,176,176}
\definecolor{darkgreen}{RGB}{0,100,0}
\definecolor{darkorange25512714}{RGB}{255,127,14}
\definecolor{forestgreen4416044}{RGB}{44,160,44}
\definecolor{lightseagreen0171189}{RGB}{0,171,189}
\definecolor{steelblue31119180}{RGB}{31,119,180}
\definecolor{darkgray176}{RGB}{176,176,176}
\definecolor{myred}{HTML}{fd7f6f}
\definecolor{mywhite}{HTML}{D8D8D8}
\definecolor{myorange}{HTML}{D7191C}
\definecolor{myblue}{HTML}{7eb0d5}
\definecolor{mygreen}{HTML}{b2e061}
\definecolor{mypurple}{HTML}{bd7ebe}
\definecolor{myorange}{HTML}{ffb55a}
\definecolor{mypurple2}{HTML}{beb9db}
\definecolor{mypink}{HTML}{fdcce5}
\definecolor{mycyan}{HTML}{8bd3c7}
\definecolor{mycyan2}{HTML}{00ffff}
\definecolor{myblue2}{HTML}{115f9a}
\definecolor{myred2}{HTML}{c23728}
\begin{axis}[
legend cell align={left},
legend columns=5,
area legend,
legend style={
  font=\LARGE,
  fill opacity=0.8,
  draw opacity=1,
  text opacity=1,
  at={(-0.06,1.08)},
  anchor=south west,
  column sep=0.2cm,
  mark size=5pt,
  draw=none
 }, 
axis lines=left,
width=\textwidth,
height =\textwidth/3.63,
tick align=center,
tick pos=left,
x grid style={darkgray176},
xmin=-2, xmax=26,
xtick style={color=black},
xtick={0,3,6,9,12,15,18,21,24},
xticklabels={Average,RM,Yelp,IMDB,DBLP,Am-p,Am-c,M-eng,M-phy},
y grid style={darkgray176},
ymin=0, ymax=1.05,
ytick style={color=black},
ytick={0.2,0.4,0.6,0.8,1.0},
yticklabel style = {font=\LARGE},
xticklabel style = {font=\LARGE}
]
\addlegendimage{ybar,ybar legend,draw=black,fill=myblue,opacity=0.9,postaction={pattern=north east lines, fill opacity=0.9},mark options={mark size=6pt,solid, line width=0.1pt},legend image code/.code={
        \draw (0cm,-0.2cm) rectangle (0.6cm,0.16cm); }} %
\addlegendentry{\qours}

\addlegendimage{ybar,ybar legend,draw=black,fill=myorange,opacity=0.9,postaction={pattern=crosshatch dots, fill opacity=0.9},mark options={mark size=6pt,solid, line width=0.1pt},legend image code/.code={
        \draw (0cm,-0.2cm) rectangle (0.6cm,0.16cm); }}
\addlegendentry{Connectivity}

\addlegendimage{ybar,ybar legend,draw=black,fill=mypink,opacity=0.9,postaction={pattern=crosshatch, fill opacity=0.9},mark options={mark size=6pt,solid, line width=0.1pt},legend image code/.code={
        \draw (0cm,-0.2cm) rectangle (0.6cm,0.16cm); }}
\addlegendentry{Eigengap}

\addlegendimage{ybar,ybar legend,draw=black,fill=mywhite,opacity=0.9,mark options={mark size=6pt,solid, line width=0.1pt},legend image code/.code={
        \draw (0cm,-0.2cm) rectangle (0.6cm,0.16cm); }}
\addlegendentry{Equal-$w$}

\addlegendimage{ybar,ybar legend,draw=black,fill=myred,opacity=0.9,postaction={pattern=north west lines, fill opacity=0.9},mark options={mark size=6pt,solid, line width=0.1pt},legend image code/.code={
        \draw (0cm,-0.2cm) rectangle (0.6cm,0.16cm); }}
\addlegendentry{Graph-Agg}

\draw[draw=black,fill=myblue,line width=0.12pt,postaction={pattern=north east lines}] (axis cs:-1.25,0) rectangle (axis cs:-0.75,0.726999083437575);

\draw[draw=black,fill=myblue,line width=0.12pt,postaction={pattern=north east lines}] (axis cs:1.75,0) rectangle (axis cs:2.25,1);
\draw[draw=black,fill=myblue,line width=0.12pt,postaction={pattern=north east lines}] (axis cs:4.75,0) rectangle (axis cs:5.25,0.929609793420046);
\draw[draw=black,fill=myblue,line width=0.12pt,postaction={pattern=north east lines}] (axis cs:7.75,0) rectangle (axis cs:8.25,0.553802816901408);
\draw[draw=black,fill=myblue,line width=0.12pt,postaction={pattern=north east lines}] (axis cs:10.75,0) rectangle (axis cs:11.25,0.929997535124476);
\draw[draw=black,fill=myblue,line width=0.12pt,postaction={pattern=north east lines}] (axis cs:13.75,0) rectangle (axis cs:14.25,0.782222222222222);
\draw[draw=black,fill=myblue,line width=0.12pt,postaction={pattern=north east lines}] (axis cs:16.75,0) rectangle (axis cs:17.25,0.604275741710297);
\draw[draw=black,fill=myblue,line width=0.12pt,postaction={pattern=north east lines}] (axis cs:19.75,0) rectangle (axis cs:20.25,0.454809177875119);
\draw[draw=black,fill=myblue,line width=0.12pt,postaction={pattern=north east lines}] (axis cs:22.75,0) rectangle (axis cs:23.25,0.561275380247035);
\draw[draw=black,fill=myorange,line width=0.12pt,postaction={pattern=crosshatch dots}] (axis cs:-0.75,0) rectangle (axis cs:-0.25,0.484773938645731);

\draw[draw=black,fill=myorange,line width=0.12pt,postaction={pattern=crosshatch dots}] (axis cs:2.25,0) rectangle (axis cs:2.75,0.989010989010989);
\draw[draw=black,fill=myorange,line width=0.12pt,postaction={pattern=crosshatch dots}] (axis cs:5.25,0) rectangle (axis cs:5.75,0.736036725325172);
\draw[draw=black,fill=myorange,line width=0.12pt,postaction={pattern=crosshatch dots}] (axis cs:8.25,0) rectangle (axis cs:8.75,0.554929577464789);
\draw[draw=black,fill=myorange,line width=0.12pt,postaction={pattern=crosshatch dots}] (axis cs:11.25,0) rectangle (axis cs:11.75,0.4239585900912);
\draw[draw=black,fill=myorange,line width=0.12pt,postaction={pattern=crosshatch dots}] (axis cs:14.25,0) rectangle (axis cs:14.75,0.322745098039216);
\draw[draw=black,fill=myorange,line width=0.12pt,postaction={pattern=crosshatch dots}] (axis cs:17.25,0) rectangle (axis cs:17.75,0.468150087260035);
\draw[draw=black,fill=myorange,line width=0.12pt,postaction={pattern=crosshatch dots}] (axis cs:20.25,0) rectangle (axis cs:20.75,0.10199547788785);
\draw[draw=black,fill=myorange,line width=0.12pt,postaction={pattern=crosshatch dots}] (axis cs:23.25,0) rectangle (axis cs:23.75,0.2813649640866);
\draw[draw=black,fill=mypink,line width=0.12pt,postaction={pattern=crosshatch}] (axis cs:-0.25,0) rectangle (axis cs:0.25,0.697532569291394);

\draw[draw=black,fill=mypink,line width=0.12pt,postaction={pattern=crosshatch}] (axis cs:2.75,0) rectangle (axis cs:3.25,0.978021978021978);
\draw[draw=black,fill=mypink,line width=0.12pt,postaction={pattern=crosshatch}] (axis cs:5.75,0) rectangle (axis cs:6.25,0.91469013006886);
\draw[draw=black,fill=mypink,line width=0.12pt,postaction={pattern=crosshatch}] (axis cs:8.75,0) rectangle (axis cs:9.25,0.377183098591549);
\draw[draw=black,fill=mypink,line width=0.12pt,postaction={pattern=crosshatch}] (axis cs:11.75,0) rectangle (axis cs:12.25,0.931476460438748);
\draw[draw=black,fill=mypink,line width=0.12pt,postaction={pattern=crosshatch}] (axis cs:14.75,0) rectangle (axis cs:15.25,0.783790849673203);
\draw[draw=black,fill=mypink,line width=0.12pt,postaction={pattern=crosshatch}] (axis cs:17.75,0) rectangle (axis cs:18.25,0.579479348458406);
\draw[draw=black,fill=mypink,line width=0.12pt,postaction={pattern=crosshatch}] (axis cs:20.75,0) rectangle (axis cs:21.25,0.451505156175207);
\draw[draw=black,fill=mypink,line width=0.12pt,postaction={pattern=crosshatch}] (axis cs:23.75,0) rectangle (axis cs:24.25,0.5641135329032);
\draw[draw=black,fill=mywhite,line width=0.12pt] (axis cs:0.25,0) rectangle (axis cs:0.75,0.636910714285714);

\draw[draw=black,fill=mywhite,line width=0.12pt] (axis cs:3.25,0) rectangle (axis cs:3.75,0.714285714285714);
\draw[draw=black,fill=mywhite,line width=0.12pt] (axis cs:6.25,0) rectangle (axis cs:6.75,0.695);
\draw[draw=black,fill=mywhite,line width=0.12pt] (axis cs:9.25,0) rectangle (axis cs:9.75,0.377);
\draw[draw=black,fill=mywhite,line width=0.12pt] (axis cs:12.25,0) rectangle (axis cs:12.75,0.923);
\draw[draw=black,fill=mywhite,line width=0.12pt] (axis cs:15.25,0) rectangle (axis cs:15.75,0.789);
\draw[draw=black,fill=mywhite,line width=0.12pt] (axis cs:18.25,0) rectangle (axis cs:18.75,0.606);
\draw[draw=black,fill=mywhite,line width=0.12pt] (axis cs:21.25,0) rectangle (axis cs:21.75,0.447);
\draw[draw=black,fill=mywhite,line width=0.12pt] (axis cs:24.25,0) rectangle (axis cs:24.75,0.544);
\draw[draw=black,fill=myred,line width=0.12pt,postaction={pattern=north west lines}] (axis cs:0.75,0) rectangle (axis cs:1.25,0.64175);

\draw[draw=black,fill=myred,line width=0.12pt,postaction={pattern=north west lines}] (axis cs:3.75,0) rectangle (axis cs:4.25,0.692);
\draw[draw=black,fill=myred,line width=0.12pt,postaction={pattern=north west lines}] (axis cs:6.75,0) rectangle (axis cs:7.25,0.844);
\draw[draw=black,fill=myred,line width=0.12pt,postaction={pattern=north west lines}] (axis cs:9.75,0) rectangle (axis cs:10.25,0.424);
\draw[draw=black,fill=myred,line width=0.12pt,postaction={pattern=north west lines}] (axis cs:12.75,0) rectangle (axis cs:13.25,0.924);
\draw[draw=black,fill=myred,line width=0.12pt,postaction={pattern=north west lines}] (axis cs:15.75,0) rectangle (axis cs:16.25,0.667);
\draw[draw=black,fill=myred,line width=0.12pt,postaction={pattern=north west lines}] (axis cs:18.75,0) rectangle (axis cs:19.25,0.593);
\draw[draw=black,fill=myred,line width=0.12pt,postaction={pattern=north west lines}] (axis cs:21.75,0) rectangle (axis cs:22.25,0.436);
\draw[draw=black,fill=myred,line width=0.12pt,postaction={pattern=north west lines}] (axis cs:24.75,0) rectangle (axis cs:25.25,0.554);
\end{axis}

\end{tikzpicture}
    }
    \vspace{-1mm}
    \caption{\revision{Clustering accuracy with alternative integrations.} }\label{fig:alternative-obj}
    \vspace{-4mm}
\end{figure}

\revision{\stitle{Alternative integrations} \label{sec:ablation-study-obj}  For the proposed spectrum-guided integration in \qours that optimizes the full objective, we compare baselines optimizing the connectivity or eigengap objective alone, setting equal weights for all view Laplacians (Equal-$w$), and directly aggregating adjacency matrices from graph views and  KNN graphs of attribute views (Graph-Agg).} Figure \ref{fig:alternative-obj} reports the clustering accuracy on each dataset and the average accuracy over all datasets.
\qours is the best in average accuracy performance and achieves the highest accuracy on almost all datasets. Optimizing connectivity or eigengap alone can achieve relatively good accuracy on some datasets but performs poorly on others, validating the design choice to combine both objectives. Despite occasional successes, assigning equal view weights often leads to poor performance, as evidenced by the low clustering quality on datasets such as RM, Yelp, and IMDB. \revision{Graph-Agg is outperformed by \qours that adopts normalized Laplacians and preserves intrinsic spectrum properties of \mvags. These results highlight the advantage of our spectrum-guided multi-view integration.}

\begin{figure}[!t]

\centering

\subfloat[\dmg]{
\resizebox{0.09\textwidth}{!}{%
\includegraphics[width=0.16\textwidth]{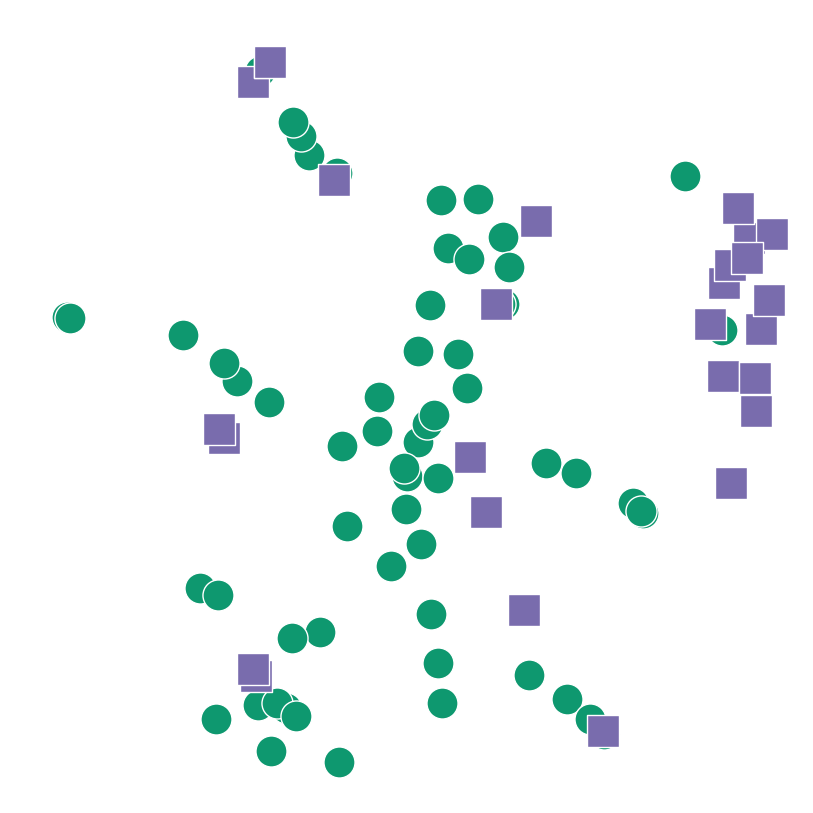}
\label{fig:tsne-rm-dmg}
 }}
\hspace{1em}
\subfloat[\pane]{
\resizebox{0.09\textwidth}{!}{%
\includegraphics[width=0.16\textwidth]{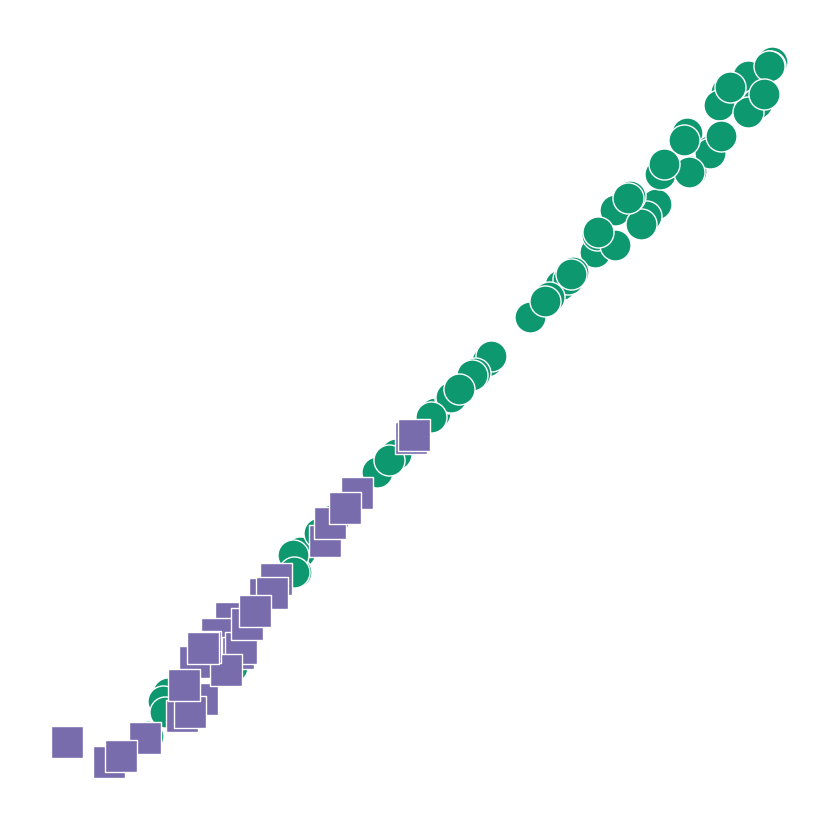}
}}
\hspace{1em}
\subfloat[\qours]{
\resizebox{0.09\textwidth}{!}{%
\includegraphics[width=0.16\textwidth]{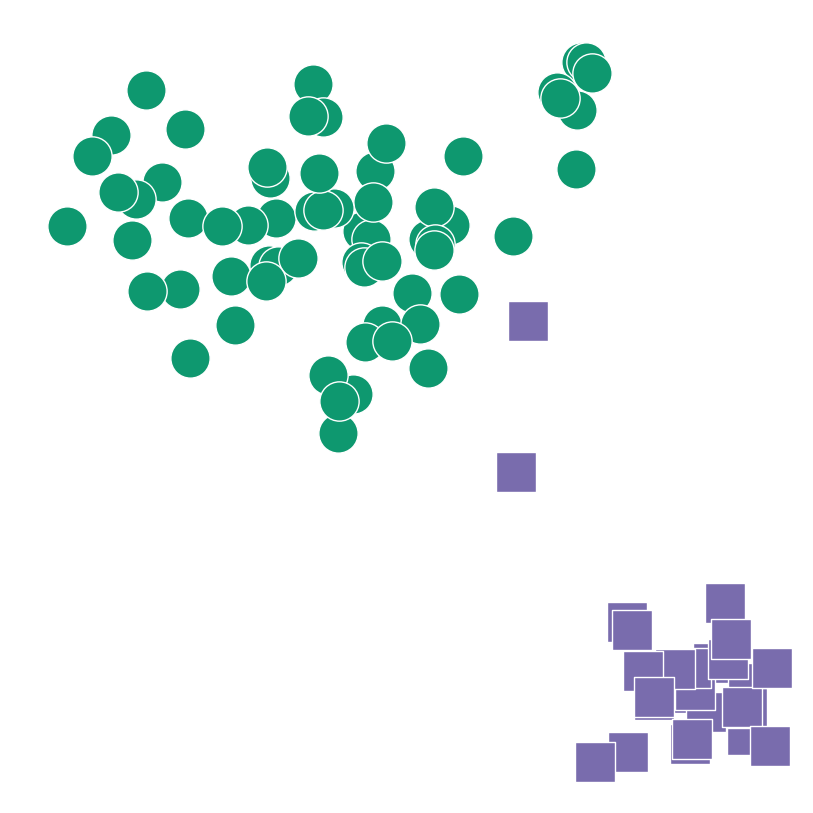}
\label{fig:tsne-rm-ours}
}}

\vspace{-2mm}

\subfloat[\conn]{
\resizebox{0.09\textwidth}{!}{%
\includegraphics[width=0.16\textwidth]{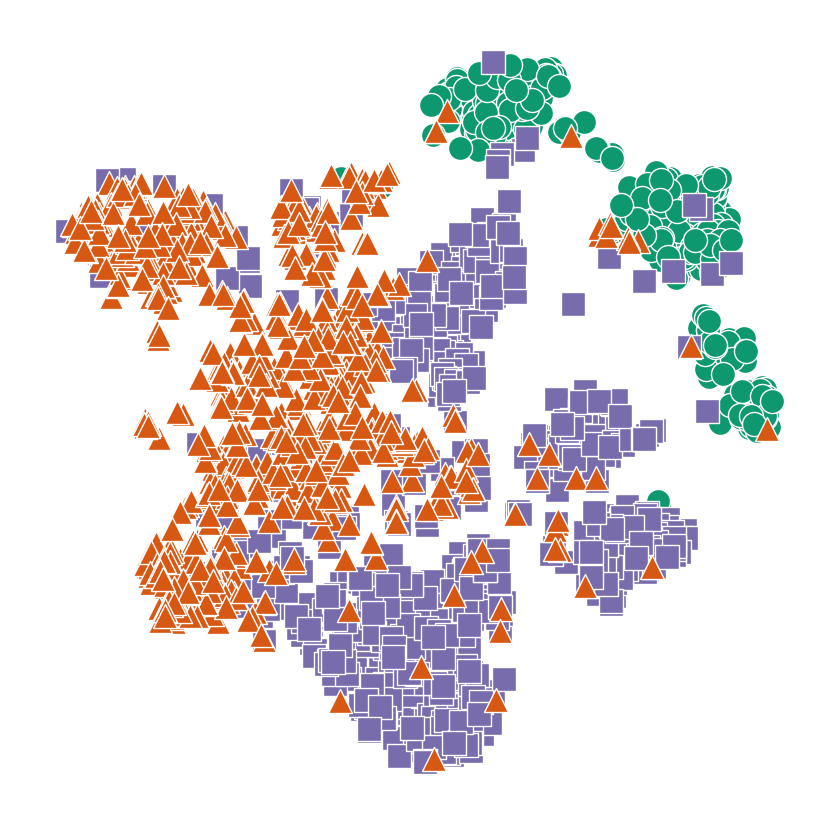}
\label{fig:tsne-yelp-conn}
}}
\hspace{1em}
\subfloat[\lmgec]{
\resizebox{0.09\textwidth}{!}{%
\includegraphics[width=0.16\textwidth]{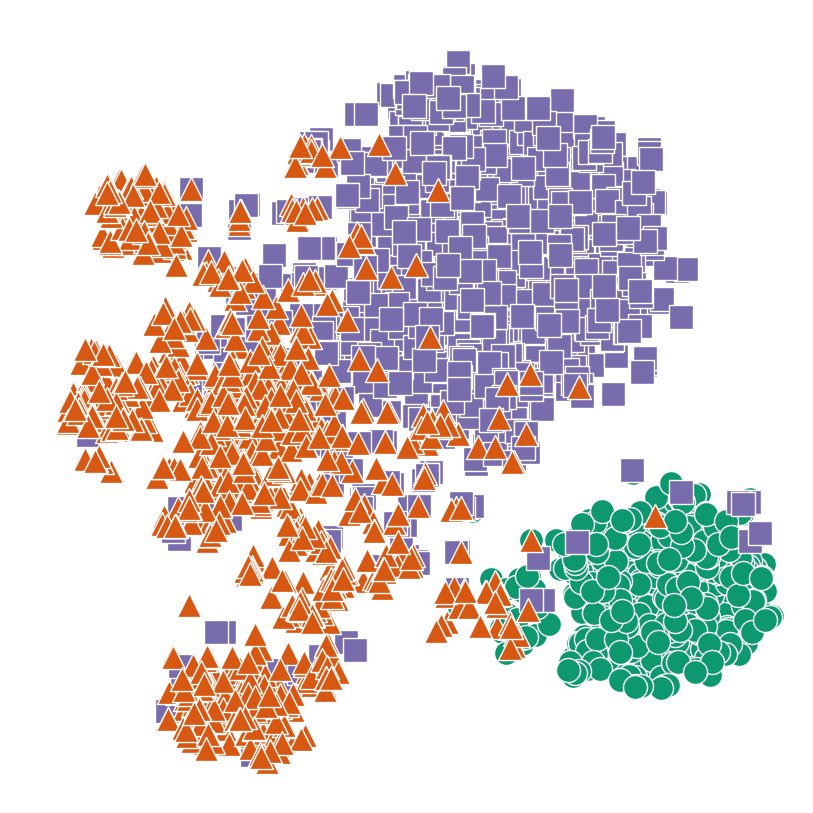}
}}
\hspace{1em}
\subfloat[\qours]{
\resizebox{0.09\textwidth}{!}{%
\includegraphics[width=0.16\textwidth]{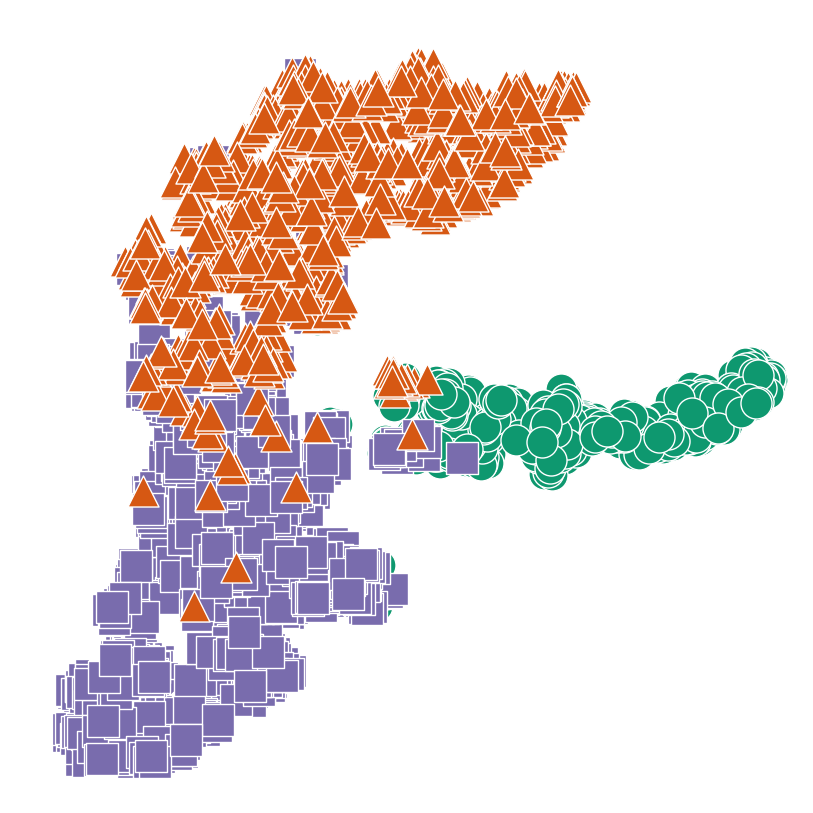}
\label{fig:tsne-yelp-ours}
}}
\vspace{-1mm}
\caption{{Embedding visualization on RM (a,b,c) and Yelp (d,e,f). Ground-truth classes are in color.}} \label{fig:tsne_fig}
\vspace{-3mm}
\end{figure}

\stitle{Embedding visualization}\label{sec:vis}
We visualize the node embeddings using t-SNE~\cite{vandermaatenVisualizingDataUsing2008} to assess the embedding quality. Due to space constraints, we present the visualizations of our method \qours and the strong baselines identified in Table \ref{tab:revise_emb} on the RM and Yelp datasets in Figure \ref{fig:tsne_fig}. On RM, shown in Figures \ref{fig:tsne-rm-dmg}-\ref{fig:tsne-rm-ours}, \qours effectively separates nodes by classes, while \dmg and \pane mix more nodes from different classes. A similar observation can be made on Yelp, shown in Figures \ref{fig:tsne-yelp-conn}-\ref{fig:tsne-yelp-ours}. The visualization demonstrates the effectiveness of our methods in generating $\LS$ for high-quality embedding.

\section{Conclusion}\label{sec:conclusion}
We propose a spectrum-guided integration scheme on \mvags with effective objectives, realized through two algorithms, \ours and \qours. 
Our approach allows classic graph algorithms to be directly applied to \mvags for clustering and embedding purposes. 
We focus on preserving community structure and node connectivity by utilizing the spectrum of Laplacian matrices.
To tackle the computational challenges, we first develop \ours with superior performance and then introduce the \qours algorithm, which is further accelerated by approximating the objective. 
Extensive experiments confirm that both \ours and \qours consistently deliver high-quality results. 
\revision{In future work, we aim to develop methods for dynamic \mvags, with 
a lazy update scheme to minimize the cost of updating view weights by executing updates only when necessary to maintain effectiveness. We will design incremental objective evaluation techniques to reduce cost. Another direction is to design  robust techniques to handle noisy MVAGs and explore GPU computation for MVAGs.}

\section*{Acknowledgment}
This work is supported by grants from the Research Grants Council of Hong Kong Special Administrative Region, China (No. PolyU 25201221, PolyU 15205224).
Jieming Shi is supported by NSFC No. 62202404, Otto Poon Charitable Foundation Smart Cities Research Institute (SCRI) P0051036-P0050643. 
This work is supported by the RGC GRF grant (No. 14217322), Hong Kong ITC ITF grant (No. MRP/071/20X), and Tencent Rhino-Bird Focused Research Grant. 

\clearpage

\bibliographystyle{IEEEtran}
\bibliography{IEEEabrv,sample}

\end{document}